\documentclass[11pt]{article}
\usepackage{amsfonts}
\usepackage{amssymb}
\usepackage{amsbsy}
\usepackage{txfonts}

\input epsf.sty

\newcommand{\BEQ}{\begin{equation}}     
\newcommand{\BEA}{\begin{eqnarray}}
\newcommand{\EEQ}{\end{equation}}       
\newcommand{\EEA}{\end{eqnarray}}
\newcommand{\eps}{\varepsilon}          
\newcommand{\del}{\delta}

\newcommand{\g}{{\mathfrak{g}}}

\newcommand{\h}{{\mathfrak{h}}}

\newcommand{\sv}{{\mathfrak{sv}}}

\newcommand{\Tr}{{\mathrm{Tr}}}

\newcommand{\Id}{{\mathrm{Id}}}

\newcommand{\SV}{{\mathrm{SV}}}
\newcommand{\Rot}{{\mathrm{Rot}}}
\newcommand{\Stab}{{\mathrm{Stab}}}

\newcommand{\eop}{$\Box$}
\newcommand{\Vect}{{\mathrm{Vect}}}
\newcommand{\diag}{{\mathrm{diag}}}

\newcommand{\vir}{{\mathfrak{vir}}}

\newcommand{\ad}{{\mathrm{ad}}}

\newcommand{\Diff}{{\mathrm{Diff}}}

\newcommand{\cotan}{{\mathrm{cotan}}}

\newcommand{\half}{{1\over 2}}
\newcommand{\Del}{\Delta}

\newcommand{\R}{\mathbb{R}}
\newcommand{\N}{\mathbb{N}}
\newcommand{\C}{\mathbb{C}}
\newcommand{\Z}{\mathbb{Z}}

\newcommand{\II}{{\rm i}}               
\renewcommand{\Re}{{\rm Re\ }}          
\renewcommand{\Im}{{\rm Im\ }}          
\renewcommand{\vec}[1]{\boldsymbol{#1}} 

\catcode`\@=11
\def\numberbysection{\@addtoreset{equation}{section}
        \def\theequation{\thesection.\arabic{equation}}}
\numberbysection

\begin{document}

\begin{titlepage}


\vskip 1.5 cm
\begin{center}
{\Large \bf A classification of periodic time-dependent generalized harmonic oscillators using a hamiltonian
action of the Schr\"odinger-Virasoro group}
\end{center}

\vskip 2.0 cm
   
\centerline{  {\bf J\'er\'emie Unterberger}$^a$}
\vskip 0.5 cm
\centerline{ B.P. 239, 
F -- 54506 Vand{\oe}uvre l\`es Nancy Cedex, France}
\vskip 0.5 cm
\centerline {$^b$Institut Elie Cartan,\footnote{Laboratoire 
associ\'e au CNRS UMR 7502} Universit\'e Henri Poincar\'e Nancy I,} 
\centerline{ B.P. 239, 
F -- 54506 Vand{\oe}uvre l\`es Nancy Cedex, France}

\begin{abstract}
\noindent 

In the wake of a preceding article \cite{RogUnt06}  introducing the Schr\"odinger-Virasoro
group, we study its affine action on a space of $(1+1)$-dimensional Schr\"odinger operators with
time- and space-dependent potential $V$  periodic in time. We focus
on the subspace  corresponding to potentials that are at most quadratic in the space
 coordinate, which is in some sense the natural quantization of the space of Hill (Sturm-Liouville) operators on the one-dimensional
torus. The orbits in this subspace  have finite codimension, and their classification
 by studying the stabilizers can be obtained by extending Kirillov's results on the
 orbits of the space of Hill operators under the Virasoro group.  We then explain the
 connection to the theory of Ermakov-Lewis  invariants
for time-dependent harmonic oscillators. These exact adiabatic invariants behave covariantly under the action
of the Schr\"odinger-Virasoro group, which allows a natural classification of the
orbits in terms of a monodromy operator on $L^2(\R)$ which is closely related to the
monodromy matrix for the corresponding Hill operator.

\end{abstract}


\vskip 1 cm

\end{titlepage}


\setcounter{section}{-1}
\section{Introduction}

The Schr\"odinger-Virasoro Lie algebra $\sv$ was originally introduced in Henkel\cite{Henk94}
as a natural infinite-dimensional extension of the Schr\"odinger algebra. Recall the latter
is defined as the algebra of projective Lie symmetries of the free Schr\"odinger equation
in (1+1)-dimensions
\BEQ (-2\II  \partial_t-\partial_r^2)\psi(t,r)=0. \label{Scheq}\EEQ
These act on equation (\ref{Scheq}) as the following first-order operators
\BEA
L_n=-t^{n+1}\partial_t-\half (n+1)t^n r\partial_r+\frac{\II }{4}
(n+1)nt^{n-1} r^2  - (n+1)\lambda t^n \nonumber \\
Y_m=-t^{m+\half}\partial_r+\II  (m+\half)t^{m-\half} r  \nonumber\\
M_p= \II  t^p \label{repsv0}
\EEA
with $\lambda=1/4$ and $n=0,\pm 1$, $m=\pm \half$, $p=0$. The $0$th-order terms in (\ref{repsv0}) correspond at the level of the group to the multiplication of the wave function
by a phase. To be explicit, the $6$-dimensional Schr\"odinger group $\cal S$ acts on $\psi$ by
the following transformations
\BEQ  (L_{-1},L_0,L_1)\ : \quad \psi(t,r)\to \psi'(t',r')=(ct+d)^{-1/2} e^{-\half\II c r^2/(ct+d)} \psi(t,r)\label{SchtrL} \EEQ
where $t'=\frac{at+b}{ct+d}$, $r'=\frac{r}{ct+d}$ with $ad-bc=1$; 
\BEQ
(Y_{\pm \half})\ : \quad \psi(t,r)\to \psi(t,r')=e^{-\II  ((vt+r_0)(r-v/2)} \psi(t,r) 
\label{SchtrY} \EEQ
where $r'=r-vt-r_0$; 
\BEQ
(M_0)\ :\quad  \psi(t,r)\to e^{-\II \gamma } \psi(t,r). \label{SchtrM} \EEQ

All together these transformations make up a group $\cal S$, called {\em Schr\"odinger group}, which is isomorphic to a semi-direct  product of $SL(2,\R)$ (corresponding to time-reparametrizations (\ref{SchtrL}))
by a Heisenberg group ${\cal H}_1$ (corresponding to the Galilei transformations
(\ref{SchtrY}), (\ref{SchtrM})). Note that the last transformation (\ref{SchtrM}) (multiplication by a constant phase) is generated by the commutators of the Galilei transformations
(\ref{SchtrY}) - these do not commute because of the added phase terms, which produce a central extension.

Now $\sv\simeq \langle L_n,Y_m,M_p\ |\ n,p\in\Z, m\in\half+\Z\rangle$ - made up of all linear combinations of the generators corresponding to all possible integer or half-integer indices - is a Lie algebra, as can be checked by direct computation. Similarly to the Lie algebra of the Schr\"odinger group, it is a semi-direct product, $\sv\simeq \g_0\ltimes \h$, where $\g_0=\langle
L_n\rangle_{n\in\Z}$ is the centerless Virasoro algebra and $\h=\langle Y_m,M_p\ |\
m\in\half+\Z,p\in\Z\}$ is a two-step nilpotent infinite-dimensional Lie algebra which
extends the Heisenberg Lie algebra. It may be exponentiated into a group (the {\it Schr\"odinger-Virasoro group}) that we denote by SV. The paper\cite{RogUnt06}, by C. Roger and the author, studies this Lie algebra for its own sake from different points of view,
including representation theory, deformations, central extensions. There is a hope that this Lie algebra or related ones
may help classify strongly anisotropic critical systems and models pertaining to out-of-equilibrium statistical physics, notably
ageing phenomena, for which the anisotropic dilation $(t,r)\to (\lambda^2 t,\lambda r)$ ($\lambda\in\R$) holds.  A systematic
investigation of the consequences  of a symmetry of the physical system under consideration  under the Schr\"odinger group or related groups has been conducted since the mid-nineties (for a short survey, see \cite{Henk05a}). 

The starting point for this work is a little different.  One of the possible motivations for introducing this Lie algebra
in the first place is that the group of Lie symmetries of {\it any} Schr\"odinger operator $-2\II \partial_t-\partial_r^2+V(t,r)$ 
may be represented as a linear combination of the generators introduced in (\ref{repsv0}). In other words, for any particular Schr\"odinger operator,
the Lie algebra of symmetries is finite-dimensional, but the symmetry algebras of all Schr\"odinger operators are contained
in $\sv$ in the above realization (see section 2.5 below for a more precise statement). The proof lies in some sense in a classical paper
by U. Niederer (see \cite{Nie74}) --  who never considered the algebra generated by all possible symmetries.

 Another related way
to look at it is that $\sv$ acts on the space of Schr\"odinger operators. More precisely, SV acts on the affine
space of Schr\"odinger operators with time- and space-dependent potential at most {\it quadratic} in the space coordinate. We 
call it ${\cal S}^{aff}_{\le 2}:=\{ -2\II \partial_t-\partial_r^2+V_2(t)r^2+V_1(t)r+V_0(t)\}$. It is assumed that
$V_0,V_1$ and $V_2$ are $2\pi$-periodic in time; this hypothesis is natural when one sets
 $t=e^{\II \theta}$ $(\theta\in\R/2\pi\Z)$ as a coordinate on the unit circle,
so that the generator $L_n$ acts as $-e^{\II n\theta}\partial_{\theta}+\ldots$. This restricted space is in some
sense minimal, which can be seen from the fact that ${\cal S}^{aff}_{\le 2}$ may be expressed in terms of three functions of time,
just like the elements of $\sv$. The phase terms in (\ref{repsv0}) add by commutation with the free Schr\"odinger equation
terms of order $1$, $r$ and $r^2$. One can show that the orbit of any Schr\"odinger operator $D\in {\cal S}^{aff}_{\le 2}$ has
finite-codimension in this space. Hence this space appears to be natural from a representation point of view.

In section 2  below (see section 2.4), we  classify the orbits of SV in ${\cal S}^{aff}_{\le 2}$. 
The classification is mainly an extension of Kirillov's results on the classification of the orbits of the space of {\it Hill 
operators}
under the Virasoro group. These are operators of the type $\partial_t^2+u(t)$. It is well-known (see for instance Guieu
\cite{Gui94} or \cite{GuiRog}) that the group of orientation-preserving diffeomorphisms $\Diff_+(\R/2\pi\Z)$ of the circle -- which exponentiates the centerless Virasoro algebra -- acts on the affine
space of Hill operators. Now the remarkable fact (despite the apparent differences between the two problems) is that
the  action of the Virasoro group $\Diff_+(\R/2\pi\Z)\subset \SV$ on the quadratic part of the potential, $V_2(t) r^2$, is equivalent to
that of $\Diff_+(\R/2\pi\Z)$ on the Hill operator $\partial_t^2+V_2(t)$.  The reason comes from the fact that the Hill operator is the
corresponding classical problem in the semi-classical limit (see section 3.2). Hence part of the classification may be borrowed directly from
the work of Kirillov (see \cite{Kir82}). A. A. Kirillov obtains his classification by studying the isotropy algebra
$Lie(\Stab_u):=\{X\in Lie(\Diff_+(\R/2\pi\Z)) \ |\ X. (\partial^2+u)=0\}.$ There is another equivalent description in terms of
the {\it lifted monodromy}, which can be explained briefly as follows. If $(\psi_1,\psi_2)$ is a basis of solutions of the
ordinary differential equation $(\partial_t^2+u(t))\psi(t)=0$, then (by Floquet's theory) 
\BEQ \left(\begin{array}{c} \psi_1(t_0+2\pi) \\ \psi_2(t_0+2\pi) \end{array}\right)= M. \left(\begin{array}{c} \psi_1(t_0) \\ 
\psi_2(t_0) \end{array}\right),\EEQ
where $M$ is some matrix (called {\it monodromy matrix}) with determinant 1 which does not depend on the base point $t_0$.
If $M$ is {\it elliptic}, i.e. conjugate to a rotation, then the eigenvectors for $M$ are multiplied by a phase $e^{\II \theta}$. If $M$ is {\it hyperbolic}, i.e.
conjugate to a Lorentz shift $\left(\begin{array}{cc} e^{\lambda} & \\ & e^{-\lambda} \end{array}\right)$, then the eigenvectors
are multiplied by a real factor $e^{\pm \lambda}$, hence the solutions of the Hill equation are unstable, going either
to zero or to infinity when $t\to \pm\infty$.  A nice
way to see it (and made rigorous in section 2) is to imagine the vector $\left(\begin{array}{c} \psi_1(t) \\ \psi_2(t) \end{array}\right)$ as 'rotating' in the plane (it may also change norm but never vanishes). The curve described by this vector may be lifted
to the Riemann surface of the logarithm for instance (obtained from the cut plane $\C\setminus\R_-$),
 so that it turns by an angle unambiguously defined
in $\R$. This gives the lifted monodromy.

The space ${\cal S}_{\le 2}^{aff}$ has been considered independently by mathematicians and physicists, with similar motivations but
different methods (that turn out to be equivalent in the end). The general idea was to solve the evolution problem associated
with $D\in {\cal S}_{\le 2}^{aff}$, i.e.to  show that the Cauchy problem $D\psi=0$ with initial condition $\psi(0,r)=\psi_0(r)$
has a unique solution and compute it explicitly. The usual method in mathematical physics for such time-dependent problems is
to consider the adiabatic approximation: if one puts formally a small coefficient $\eps$ in front of $\partial_t$, the problem
is equivalent by dilating the time coordinate to the equation $(-2\II \partial_t-\partial_r^2+V(\eps t,r))\psi=0$, so
that $V$ is a potential that is slowly varying in time. Suppose that $\Del_{\eps}(t):=-\partial_r^2+V(\eps t,r)$ has a pure point spectrum
$\{\lambda_n(t), n\in\N\}$  for
every $t$, where $\lambda_n$ is $C^{\infty}$ in $t$, say,   and let $\psi_n(t)$ be a normalized 
eigenfunction of $\Del_{\eps}(t)$ satisfying
the {\it gauge-fixing condition} $\langle \psi_n(t), \dot{\psi}_n(t)\rangle=0$. Then there exists a {\it parallel transport
operator} $W(s,t)$ carrying the eigenspace with eigenvalue $\lambda_n(s)$ to the eigenspace with
eigenvalue $\lambda_n(t)$, and a {\it phase operator} $\Phi(s,t)$, given simply by the multiplication by a phase $e^{\frac{\II}{2}
\int^t \lambda_n(s)\ ds}$ on each
eigenspace, such that the solution of the Schr\"odinger equation is given at first order in $\eps$ by the composition of $W$ and $\Phi$.  One may see the solutions formally as flat sections for a connection (called {\it Berry connection}) related in simple
terms
to the phase operator (see \cite{BohMosKoiNiuZwa}). This scheme may be iterated, giving approximate solutions to the Schr\"odinger equation that are
correct to any order in $\eps$ (see for instance Joye \cite{Joy92}), but it is rarely the case that one can give {\it exact} solutions.  By considering
the related classical problem, G. Hagedorn (see \cite{Hag98})
constructs a set of raising and lowering operators (generalizing those associated to the usual harmonic oscillator)
 for general Schr\"odinger operators
in ${\cal S}_{\le 2}^{aff}$, and uses them to  solve the equation explicitly. The same set of operators had been considered
previously by two quantum physicists, H. R. Lewis and W. B. Riesenfeld (see \cite{LewRie}), and obtained by looking for
an exact invariant, i.e. for a time-dependent operator $I(t)$ (not including the time-derivative) such that
$\frac{dI}{dt}=\frac{\partial I}{\partial t}+\II [I(t),\half(\partial_r^2-V(t,r))]=0$. They find for each operator $D$ in ${\cal S}_{\le 2}^{aff}$ a family of invariants (called sometimes the {\it Ermakov-Lewis invariants}, see \cite{Pad00})  depending on an 
arbitrary real solution $\xi$ of a certain differential equation of order 3 (see Proposition 3.1.4),
constructed out of generalized raising and lowering operators and spectrally equivalent to the
 standard harmonic oscillator $-\half(\partial_r^2-r^2)$. These invariants have been used to solve quite a few physical problems, ranging from quantum mechanics for charged 
particles to cosmology (see \cite{GuaMoy03}, \cite{GuaMoy03b}, \cite{Pad00}, \cite{RayRei82} for instance). It turns out that very few Schr\"odinger operators have an exact invariant of the type $I(t)=f_2(t,r)\partial_r^2+f_1(t,r)\partial_r+f_0(t,r)$. These may be expressed, as shown
by H. Lewis and P. Leach (see \cite{LeaLew}), in terms of three arbitrary functions of time (the exact expression is complicated).
Exact invariants allow in principle
  to solve explicitly the original problem, at least if one knows how to diagonalize them (which is the case here).
Hence (provided one requires that an exact invariant exists) the space ${\cal S}_{\le 2}^{aff}$ is maximal.

There are three new features here:

-- {\em the action of the Schr\"odinger-Virasoro group} on ${\cal S}_{\le 2}^{aff}$ (which is essentially a conjugate action, leaving
all invariant quantities unchanged, for instance the spectrum and the monodromy) makes it possible to reduce the study to five
families of operators, with qualitatively different properties (see section 2.4). They are mainly
characterized by the monodromy of the associated Hill operator $\partial_t^2+V_2(t)$, but there also appear some non-generic
orbits in cases when the quadratic and linear parts of the potential are 'resonant'. The non-periodic case is much simpler, since ({\em locally}
in time) all Schr\"odinger operators in ${\cal S}_{\le 2}^{aff}$ are formally equivalent (see section 3 below). The coefficients of the Ermakov-Lewis invariants
are related in a very simple way to the invariants of the orbits;

-- one is interested in Schr\"odinger operators with {\it time-periodic potential}. Hence one may consider (as in the case of ordinary differential operators, see above) the monodromy, which is a bounded operator acting on $L^2(\R)$. The
monodromy operator is given explicitly  and shown to be closely related to the classical monodromy of the related Hill operator;

-- the computation of the monodromy in the case when the associated Hill operator is {\it hyperbolic}(see above) requires the use
 of an Ermakov- Lewis invariant associated to a {\it purely imaginary} function $\xi$, which is equivalent to the standard harmonic 'repulsor'
 $-\half(\partial_r^2+r^2)$. The reason (explained more precisely in section 3 below)
is that the usual Ermakov-Lewis invariants are defined only if  $I_{V_2}(\xi)>0$, where the  invariant quantity $I_{V_2}(\xi)$ ({\em quadratic}
in $\xi$) is
associated to the Hill operator $\partial_t^2+V_2(t)$
and its stabilizer $\xi(t)$ in Lie($\Diff_+(\R/2\pi\Z)$). The stabilizer satisfies a linear differential equation of order 3 and
has generically only one periodic solution (up to a constant). If one does not require $\xi$
to be {\it periodic}, then $I_u(\xi)$ may be chosen to be positive, which is perfectly suitable for a local study (in time) but  is
of little practical use for  the computation of the monodromy. If however one requires that $\xi$ be periodic, then $I_u(\xi)$ is {\it negative}
in the hyperbolic case, unless one chooses $\xi$ to be purely imaginary. Hence one is naturally led to use the spectral decomposition
of the harmonic 'repulsor' (which has an absolutely continuous spectrum equal to the whole real line). Usually there is no adiabatic
scheme, hence no phase operator, in the case when eigenvalues are not separated by a gap. But in this very particular case, such
a phase operator may be computed and is very analogous to that obtained in the elliptic case, for which the spectrum is discrete.
There exists also some non-generic cases (corresponding to a unipotent monodromy matrix for the underlying Hill operator)
for which $I_{V_2}(\xi)=0$. The natural invariant is then spectrally equivalent either to  the bare Laplacian $-\half\partial_r^2$ or
to the Airy operator $-\half(\partial_r^2-r)$.

One of the main results may be stated as follows (see sections 3.4,3.5,3.6): the monodromy operator is unitarily equivalent to the
unitary multiplication operator $f(k)\to e^{\II kT-\II\pi\gamma} f(k)$, where $\gamma$ is some constant and 
$k$ is the spectral parameter of the {\em model operator}
$-\half(\partial_r^2+\kappa r^2)$ ($\kappa=\pm 1,0$) or $-\half(\partial_r^2-r)$,
 and $T=\int_0^{2\pi} \frac{du}{\xi(u)}$ ($\xi$ real) or $\II \int_0^{2\pi}
\frac{du}{\xi(u)}$ ($\xi$ imaginary). The above integrals must be understood in a generalized sense if $\xi$ has some zeros;
a complex deformation of contour is needed then. Comparing with the usual Berry phase $e^{\frac{\II}{2} \int_0^{2\pi} \lambda_k(s)\ ds}$,
one sees that the eigenvalue $\lambda_k(t)=-2k$ is {\it constant}, but that the natural (possibly singular) time-scale is
$\tau:=\int^t \frac{du}{\xi(u)}$.

The paper is organized as follows.

 Section 1 is preliminary and contains notations and  results (contained in \cite{RogUnt06})
concerning the Schr\"odinger-Virasoro group and its action on Schr\"odinger operators.

 Section 2 is dedicated to the classification
of the orbits and of the isotropy subgroups $G_D:=\{g\in\ SV\ |\ g.D=D\}$, $D\in {\cal S}_{\le 2}^{aff}$ (see section 2.4).
It  contains long but necessary preliminaries on the action of the Virasoro group on Hill operators. The connection
to the results of U. Niederer is made in the last paragraph.

 We solve the monodromy problem for the Schr\"odinger operators
of the form $-2\II \partial_t-\partial_r^2+V_2(t)r^2+\gamma$  ($\gamma$ constant)
in section 3. We study first  the corresponding 
classical problem  given by the associated Hill operator, $\ddot{x}+V_2(t)x=0$ (an ordinary differential equation).
 The solution of the quantum problem is then easily deduced from that of
the classical problem. In either case, the monodromy is obtained
by relating  the Ermakov-Lewis invariants  to the orbit data.

Finally, we show in section 4 how to parametrize a general Schr\"odinger operator $-2\II \partial_t-\partial_r^2+V_2(t)r^2
+V_1(t) r+V_0(t)\in {\cal S}_{\le 2}^{aff}$ by means of a
{\it three-dimensional invariant} $(\xi(t),\del_1(t),\del_2(t))$ (see Definition 4.2). The parametrization is one-to-one
or 'almost' one-to-one depending on the orbit class of the potential $V_2$ ($SV$-orbits in ${\cal S}_{\le 2}^{aff}$ have generically
codimension 2, whereas adjoint orbits corresponding to the invariant have generically codimension 2 or 3).
 The action of the Schr\"odinger-Virasoro group on ${\cal S}_{\le 2}^{aff}$, once written in terms of the invariant, becomes much simpler, and is easily shown to be  Hamiltonian for a natural
symplectic structure. A generalized Ermakov-Lewis invariant may also be written in terms of this three-dimensional invariant.
 We then solve the monodromy
cases for the 'resonant' cases left from section 3.

{\em Notation:} The notation $\Vect(S^1):=\{\xi(t)\partial_t \ |\ \xi\in\C^{\infty}(\R/2\pi\Z)\}$ will be used for
the Lie algebra of $C^{\infty}$-vector fields on the torus $\R/2\pi\Z$. The infinite-dimensional group $\Diff_+(\R/2\pi\Z)$ of orientation-preserving diffeomorphisms of the torus  $\R/2\pi\Z$ (also called {\it centerless Virasoro group})  has a Lie structure, and its Lie algebra is
$\Vect(S^1)$ (see \cite{GuiRog} for details).

Let us gather here (for the convenience of the reader) a few notations scattered in the text. Time and space coordinates are usually (at least starting from section 2) denoted by $\theta$ and $x$ (see explanations before Lemma 1.6
for the passage to Laurent coordinates $(t,r)$).  Stabilizers
in $\Vect(S^1)$ of the Hill operator $\partial_{\theta}^2+V_2(\theta)$ are usually denoted by $\xi$ 
(which is either real or
purely imaginary). If $\xi$ is purely imaginary, then one sets $\xi:=\II\eta$. As for (operator) invariants of
the Schr\"odinger operators (see section 4), we write them as $\half\left[a(\theta)x^2-b(\theta)\partial_x^2-\II c(\theta)
(x\partial_x+\partial_x x)+d(\theta)(-\II\partial_x)+e(\theta)x+f(\theta)\right].$ The correspondence between
the vector invariant $(\xi,\del_1,\del_2)$ and the operator (generalized Ermakov-Lewis) invariant is given in
Theorem 4.4. 


\section{The Schr\"odinger-Virasoro group and its action on Schr\"odinger operators}


We recall in this preliminary section the properties of the Schr\"odinger group proved
in \cite{RogUnt06} that will be needed throughout the article.

{\bf Definition 1.1} (see \cite{RogUnt06}, Definition 1.2)

{\em 
 We denote by $\sv(\kappa)$, $\kappa=0$ or $\half$,  the Lie algebra with generators
$
L_n,Y_m,M_n (n\in\Z,m\in\kappa+\Z)$ and
  following relations (where $n,p\in\Z,m,m'\in\kappa+\Z$) :
$$
[L_n,L_p]=(n-p)L_{n+p}
$$
$$
[L_n,Y_m]=({n\over 2}-m)Y_{n+m},\quad [L_n,M_p]=-pM_{n+p};
$$
$$
[Y_m,Y_{m'}]=(m-m')M_{m+m'},
$$
$$  [Y_m,M_p]=0,\quad  [M_n,M_p]=0.
$$

If $f$ (resp. $g$, $h$) is a Laurent series, $f=\sum_{n\in \Z} f_n t^{n+1}$, resp. $g=\sum_{n\in\kappa+\Z} g_n t^{n+\half}$,
$h=\sum_{n\in\Z}  h_n t^n$, then we shall write
\BEQ {\cal L}_f=\sum f_n L_n,\quad {\cal Y}_g=\sum g_n Y_n,\quad {\cal M}_h=\sum h_n M_n.\EEQ
}

Note the shift in the indices in the Laurent series which disappears in the Fourier coordinates, see remarks preceding Lemma 1.6
below.

It is often unimportant (or a matter of taste) in this paper whether the shift index $\kappa$ is $0$ or $\half$
 (see remarks after Theorem 2.4.2 though). In this section (unless otherwise stated) $\sv$ stands indifferently
 for $\sv(0)$ or $\sv(1/2)$. In the following sections,
we shall abbreviate $\sv(0)$ to $\sv$ for convenience.

{\bf Definition 1.2} (see \cite{RogUnt06}, Definition 1.3)

{\em
  Denote by $d\pi_{\lambda}$ the representation
 of $\sv$ as differential operators of order one
on $\R^2$ with coordinates $t,r$ defined by
\BEA
d\pi_{\lambda}(L_n)=-t^{n+1}\partial_t-\half (n+1)t^n r\partial_r+\frac{1}{4}\II
(n+1)nt^{n-1} r^2  - (n+1)\lambda t^n \nonumber \\
d\pi_{\lambda}(Y_m)=-t^{m+\half}\partial_r+\II (m+\half)t^{m-\half} r  \nonumber\\
d\pi_{\lambda}(M_p)=\II  t^p  \label{repsv}
\EEA
}

{\bf Proposition 1.3}  (see \cite{RogUnt06}, Theorem 1.1)

{\em
\begin{enumerate}
\item
The Lie algebra $\sv$ can be exponentiated to a Lie group denoted by $SV$. It is isomorphic
to a semi-direct product $SV=G_0\ltimes H$, where $G_0\simeq \Diff_+(\R/2\pi\Z)$ is the group of orientation-preserving diffeomorphisms of the torus $\R/2\pi\Z$, and
$H\simeq C^{\infty}(\R/2\pi\Z)\times C^{\infty}(\R/2\pi\Z) $ (as a vector space) is  the product of two copies
of the space of
infinitely differentiable functions on the circle, with its group
structure modified as follows:
\BEQ
(\alpha_2,\beta_2).(\alpha_1,\beta_1)=(\alpha_1+\alpha_2,\beta_1+\beta_2+\half(
{
\alpha}'_1
\alpha_2-\alpha_1\alpha'_2)).
\EEQ

The semi-direct product is given by:
\BEQ
(1;(\alpha,\beta)).(\phi;0)=(\phi;(\alpha,\beta))
\EEQ
and
\BEQ
(\phi;0).(1;(\alpha,\beta))=(\phi;
((\phi')^{\half}(\alpha\circ\phi),\beta
\circ\phi)).
\EEQ

\item
The infinitesimal representation $d\pi_{\lambda}$ of $\sv$ can be exponentiated to 
the following  representation of the group $SV$ on $C^{\infty}$ functions of two
variables,
\begin{itemize}
\item[(i)] $$(\pi_{\lambda}(\phi;0)f)(t',r')=(\phi'(t))^{-\lambda} e^{\frac{1}{4}\II
 \frac{\phi''(t)}{\phi'(t)} r^2} f(t,r) $$
if $\phi\in \Diff_+(\R/2\pi\Z)$ induces the coordinate change $(t,r)\to (t',r')=(\phi(t),
r\sqrt{\phi'(t)})$;
\item[(ii)] $$(\pi_{\lambda}(1;(\alpha,\beta))f)(t',r')=e^{-\II 
(\alpha'(t)r-\half \alpha(t)\alpha'(t)+\beta(t))} f(t,r) $$
if $(\alpha,\beta)\in C^{\infty}(\R/2\pi\Z)\times C^{\infty}(\R/2\pi\Z)$ induces
the coordinate change $(t,r)\to (t,r')=(t,r-\alpha(t)).$
\end{itemize}
\end{enumerate}
}

{\bf Definition 1.4} (see \cite{RogUnt06}, Definition 2.1)

{\em
 Let ${\cal S}^{lin}$ be the vector space of second order
operators on $\R^2$ defined by
$$D\in{\cal S}^{lin}\Leftrightarrow D=h(-2\II\partial_t-\partial_r^2)
+V(t,r),
\quad h,V\in C^{\infty}(\R^2)$$
and ${\cal S}^{aff}\subset {\cal S}^{lin}$ the affine subspace of
'Schr\"odinger operators' given by the hyperplane $h=1$.

In other words, an element of ${\cal S}^{aff}$ is the sum of the free
Schr\"odinger
operator
$-2\II\partial_t-\partial_r^2$ and of a potential $V$.
}

{\bf Proposition 1.5} (see \cite{RogUnt06}, Proposition 2.5, Proposition 2.6)

{\em
Let $\sigma_{1/4}: SV\to Hom({\cal S}^{lin},{\cal S}^{lin})$ the representation
of the group of SV on the space of Schr\"odinger operators defined by the left-and-right
action
$$\sigma_{1/4}(g): D\to \pi_{5/4}(g) D \pi_{1/4}(g)^{-1},\quad
g\in SV, D\in {\cal S}^{lin}.$$
Then $\sigma_{1/4}$ restricts to an affine action on the affine subspace ${\cal S}^{aff}$ which is given by the following formulas:
\BEA
\sigma_{1/4}(\phi;0).(-2\II \partial_t-\partial_r^2+V(t,r))=\nonumber\\
\quad 
-2\II  \partial_t-\partial_r^2+\phi'(t) V(\phi(t),r\sqrt{\phi'(t)})+\half  r^2 \Theta(\phi)(t)  \\
\sigma_{1/4}(1;(a,b)).(-2\II \partial_t-\partial_r^2+V(t,r))=\nonumber \\
\quad -2\II \partial_t-\partial_r^2+V(t,r-a(t))-2 ra''(t)- (2b'(t)-a(t)a''(t)).
\EEA

where $\Theta:\phi\to \frac{\phi'''}{\phi'}-\frac{3}{2} \left(\frac{\phi''}{\phi'}\right)^2$ is the Schwarzian derivative. 
}

One may also consider a generalized left-and-right action $\sigma_{\lambda}(g): D\to \pi_{\lambda+1}(g)D
\pi_{\lambda}(g)^{-1}$, but then the subspace ${\cal S}_2^{aff}$ (see Definition 2.1.2) is not preserved by
$\sigma_{\lambda}|_{\Diff_+(\R/2\pi\Z)}$ any more, which ruins all subsequent computations. Actually $1/4$ corresponds
to the 'scaling dimension' of the Schr\"odingerian field in one dimension (see \cite{HenUnt03}).

\vskip 1 cm

We shall occasionally use the time-reparametrization
\BEQ \phi: \R/2\pi\Z \to S^1\simeq U(1) ,\quad \theta\to t=e^{\II \theta} \EEQ
from the torus to the unit circle. It allows to switch from the Fourier coordinate
$\theta$ to the Laurent coordinate $t$. In particular, 
\BEQ {\cal L}_{t^{n+1}}=\pi_{1/4}(\phi;0) \   {\cal L}_{e^{\II n\theta}}\ \pi_{1/4}(\phi;0)^{-1},\quad
{\cal Y}_{t^{n+\half}}=\pi_{1/4}(\phi;0) \   {\cal Y}_{e^{\II n\theta}}\ \pi_{1/4}(\phi;0)^{-1},\quad
{\cal M}_{t^n}=\pi_{1/4}(\phi;0) \   {\cal M}_{e^{\II n\theta}}\ \pi_{1/4}(\phi;0)^{-1}.
\EEQ
If $n$ is an integer, ${\cal Y}_{t^{n+\half}}$ should be understood to be acting on the two-fold covering
of the complex plane where the square-root is defined; conversely, if $n$ is a half-integer, then
${\cal Y}_{e^{\II n\theta}}$ acts on  $4\pi$-periodic functions. In other words, the 'natural' choice for $\sv$
should be $\sv(\half)$, resp. $\sv(0)$ in the Laurent, resp. Fourier coordinates.

 Applying formally the formulas of Proposition 1.3,
one gets
\BEQ (\pi_{1/4}(\phi;0)^{-1} f)(\theta,x)=(\II e^{-\II \theta})^{1/4} e^{-\frac{1}{4}
\II  x^2} f(e^{\II\theta}, \pm xe^{\II(\frac{\theta}{2}+\frac{\pi}{4})} ) \EEQ
(with some ambiguity in the sign) which is an $8\pi$-periodic function. Applying
now (still formally) Proposition 1.5 yields the following result, which can be
checked by direct computation.

{\bf Lemma 1.6}

{\em
Let $f(t,r)$ be a solution of the Schr\"odinger equation
$$(-2\II \partial_t-\partial_r^2+V(t,r))f(t,r)=0.$$
Then 
\BEQ \tilde{f}:(\theta,x)\to e^{-\II \theta/4} e^{-\frac{1}{4} \II x^2}
f(e^{\II \theta},xe^{\II(\theta/2+\pi/4)}) \EEQ
is a solution of the transformed Schr\"odinger equation
\BEQ
\left[-2\II\partial_{\theta}-\partial_x^2+\frac{1}{4} x^2+\II e^{\II \theta}
V(e^{\II\theta},xe^{\II(\theta/2+\pi/4)}) \right] \tilde{f}(\theta,x)=0.
\EEQ
}

{\it In the following sections, we shall (except when explicitly mentioned) always work with the Lie algebra
$\sv(0)$ in the Fourier coordinates $\theta,x$  (i.e. the Lie algebra 
generated by the ${\cal L}_f$, ${\cal Y}_g$ and ${\cal M}_h$ with
$2\pi$-periodic functions $f,g,h$), and write $\sv$ instead of $\sv(0)$ for simplicity.}


\section{Classification of the Schr\"odinger operators in ${\cal S}_{\le 2}^{aff}$}

From now on, whe shall concentrate on the affine subspace of Schr\"odinger operator with potentials which are at most quadratic
in the space coordinate. As mentioned in the Introduction, this subspace is invariant under the action of SV. The
purpose of this section is to classify the orbits.

\subsection{Statement of the problem and connection with the classification of Hill operators}

Let us first define  two natural subspaces of ${\cal S}^{aff}$.

{\bf Definition 2.1.1 (Schr\"odinger operators with at most quadratic potential)} (see \cite{RogUnt06}, Prop. 2.6) 

{\em
Let ${\cal S}^{aff}_{\le 2}=\{-2\II\partial_{\theta}-\partial_x^2+V_2(\theta)x^2+V_1(\theta)x+V_0(\theta)\}\subset {\cal S}^{aff}$ be the affine space of Schr\"odinger operators with a potential which is $2\pi$-periodic in time and at most 
quadratic  in the coordinate $x$.
}

{\bf Definition 2.1.2 (Schr\"odinger operators with quadratic potential)}

{\em
Let ${\cal S}^{aff}_{2}=\{-2\II\partial_{\theta}-\partial_x^2+V_2(\theta)x^2\}\subset {\cal S}^{aff}_{\le 2}$ be the affine space of
  Schr\"odinger operators in ${\cal S}^{aff}_{\le 2}$  with
time-periodic  potential proportional to $x^2$.
}

We do not assume $V_2$ to be positive. Hence what we really consider are harmonic
'oscillators-repulsors', corresponding to the quantization of a classical oscillator-repulsor with time-dependent Hamiltonian
$\half( p^2+V_2(\theta)x^2+V_1(\theta)x+V_0(\theta))$. If $V_1\equiv 0$, then t
he classical equation of motion $\frac{d^2 x}{d\theta^2}=-V_2(\theta)x-\half V_1(\theta)$ has $0$ as an attractive, resp. repulsive
fixed point depending on the sign of $V_2$. If $V_2$ is not of constant sign, things can be  complicated; it is  not clear a priori
whether solutions are stable or unstable. We shall come back to this problem (which turns out to be more or less equivalent
to the a priori harder quantum problem, at least as far as monodromy in concerned) in section 3.2.

The first subspace ${\cal S}^{aff}_{\le 2}$ is preserved by the action of SV (see Proposition 1.5) 
and is in some sense minimal (the SV-orbit of  the free Schr\"odinger equation, or
of the standard harmonic oscillator $-2\II\partial_{\theta}-\partial_x^2+a^2 x^2$, contains
'almost' all potentials which are at most  quadratic in $x$). As we shall prove below, the orbits
in ${\cal S}^{aff}_{\le 2}$ have finite codimension. 

Let us write down for the convenience of the reader the restriction of the action of $\sigma_{1/4}$ to ${\cal S}_{\le 2}^{aff}$:
let $D=-2\II \partial_{\theta}-\partial_x^2+V_2(\theta)x^2+V_1(\theta)x+V_0(\theta)$, then
\BEQ \sigma_{1/4}(\phi;0)(D)=-2\II \partial_{\theta}-\partial_x^2+\left( \phi'^2 \ .\ V_2\circ\phi+\half \Theta(\phi)\right)x^2+
\left(\phi'^{3/2}\ .\  V_1\circ\phi\right) x+\phi' \ .\ V_0\circ\phi  \label{actphi} \EEQ
-- recall $\Theta(\phi)=\frac{\phi'''}{\phi'}-\frac{3}{2} \left(\frac{\phi''}{\phi'}\right)^2$ is the Schwarzian derivative --, and
\BEQ \sigma_{1/4}(1;(a,b))(D)=-2\II \partial_{\theta}-\partial_x^2+V_2 x^2 +(V_1-2aV_2-2a'')x+(V_0-aV_1+a^2 V_2-2b'+aa'') \label{actab},\EEQ
while the infinitesimal action is given by
\BEQ d\sigma_{1/4}({\cal L}_f)(D)=-(\half f'''+2f'V_2+fV'_2)x^2-(fV'_1+\frac{3}{2} f'V_1)x-(fV'_0+f'V_0),
\label{actdphi} \EEQ
\BEQ d\sigma_{1/4}({\cal Y}_g+{\cal M}_h)(D)=-2(g''+gV_2)x-(2h'+gV_1). \label{actdab} \EEQ

These four formulas are fundamental for most computations below, and we shall constantly refer to them.

Similarly, ${\cal S}^{aff}_{2}$ is preserved by the $\sigma_{1/4}$-action of $\Diff_+(\R/2\pi\Z)$ (see
Proposition 1.5). It turns out that the orbit
theory for this space is equivalent to that of the Hill operators under the Virasoro group.
Let us first give some notations and recall basic facts concerning Hill operators.

{\bf Definition 2.1.3}

{\em
A Hill operator is a Sturm-Liouville operator on the one-dimensional torus, i.e.
 a second-order operator of the form $\partial_{\theta}^2+u(\theta)$ where
$u(\theta)\in C^{\infty}(\R/2\pi\Z)$ is a $2\pi$-periodic function.
}

The action of the  group of time-reparametrizations  on a Hill operator may be constructed as follows. Starting 'naively' from the simple action of diffeomorphisms on functions,

$$ \psi\to \psi\circ \phi, \quad \phi\in \Diff_+(\R/2\pi\Z),$$
one sees that $(\partial^2+u)(\psi)=0$ is equivalent to the transformed equation 
$(\partial^2+p(\theta)\partial+q(\theta))(\psi\circ\phi)=0$ if one sets
$p=-\frac{\phi''}{\phi'}$ and $q=\phi'^2 \ .\ u\circ\phi$. Then one uses the following:

{\bf Definition 2.1.4 (Wilczinsky's semi-canonical form)} (see Magnus-Winkler, \cite{MagWin}, 3.1, or
Guieu, \cite{Gui94}, Proposition 2.1.1)

{\em

If $\psi$ is a solution of the second-order equation $(\partial^2+p(\theta)\partial+q(\theta))\psi=0$,
then $\tilde{\psi}:=\lambda(\theta)\psi$ is a solution of the Hill equation $(\partial^2+u(\theta))\tilde{\psi}=0$ provided
\BEQ \lambda(\theta)=\exp\left(\half \int_{\theta_0}^{\theta} p(s)\ ds\right) \EEQ
for some $\theta_0$ and
\BEQ u=-\half p'-\frac{1}{4} p^2+q.\EEQ
}

One obtains in this case $\lambda=(\phi')^{-1/2}$, and the transformed operator reads:
$\partial^2+(\phi')^{2}\ .\  u\circ\phi + \half \Theta(\phi)$, where $\Theta$ is the
Schwarzian derivative. The presence of this last term shows that this transformation defines a {\it
projective} action of $\Diff_+(\R/2\pi\Z)$. Summarizing, one obtains:

{\bf Proposition 2.1.5} (see Guieu, \cite{Gui94} or Guieu-Roger, \cite{GuiRog})

{\em
The transformation 
$$\partial^2+u\to \phi_*(\partial^2+ u):=\partial^2+(\phi')^2 \ .\ u\circ\phi +\half \Theta(\phi)$$
defines an action of $\Diff_+(\R/2\pi\Z)$ on the space of Hill operators, which is equivalent to the affine
 coadjoint action on $\vir^*_{\half}$ (i.e. with central charge $c=\half$). A solution of the transformed equation
may be obtained from a solution $\psi$ of the initial equation $(\partial^2+u)\psi=0$
by setting $\phi_* \psi=(\phi')^{-\half} \psi\circ\phi.$ In other words, the solutions
of the Hill equations behave as $(-\half)$-densities.
}

\bigskip

The important remark now is the following:

{\bf Lemma 2.1.6}

{\em
The above action of $\Diff_+(\R/2\pi\Z)$ on the space of Hill operators is equivalent to
 the $\sigma_{1/4}$-action of\  $\Diff_+(\R/2\pi\Z)$ on the space ${\cal S}_2^{aff}$.
}

 Namely, Proposition 1.5 above (see also (\ref{actphi}))  shows
that
\BEQ \sigma_{1/4}(\phi) (-2\II\partial_{\theta}-\partial_x^2+V_2(\theta)x^2)=
-2\II \partial_{\theta}-\partial_x^2 + \tilde{V}_2(\theta)x^2, \EEQ
where the potential $\tilde{V}_2$ is the image of $V_2$ (viewed as the potential of a  Hill operator in the
coordinate $\theta$) by the diffeomorphism $\phi$, i.e. $\phi_*(\partial_{\theta}^2+V_2(\theta))=
\partial_{\theta}^2+\tilde{V}_2(\theta)$. Once again, this should not come as a surprise since the Hill equation is the
semi-classical limit of the Schr\"odinger operator (see section 3.2). \hfill \eop

So we shall need to recall briefly the classification of the orbits of Hill operators under the Virasoro
group. There are mainly three a priori different classifications, which of course turn out in the end to be
equivalent: the first one is by the {\em lifted monodromy} of the solutions (see for instance
B. Khesin and R. Wendt, \cite{KheWen}); the second one consists in looking for normal forms for the solutions,
either an exponential form for non-vanishing solutions or a standard form for a dynamical system associated
with the repartition of the zeros (see the article by V. F. Lazutkin and T. F. Pankratova,  \cite{LazPan75}); the third one, due to A. A. Kirillov (see \cite{Kir82})  proceeds in a more indirect way by looking at the
isotropy groups. We shall need the first and the last classification for our purposes. They are the
subject of the two upcoming subsections (see also \cite{BalFehPal} for a related review and 
application to the global Liouville equation)..


\subsection{Classification of Hill operators by the lifted monodromy}


Let us now turn to the  classification of the orbits under the Virasoro group 
of the space of Hill operators. 

Consider a pair $(\psi_1,\psi_2)$ of linearly independent solutions of the Hill equation
$(\partial^2+u)\psi=0$.  It is a classical result (a particular case of Floquet's theory
for Schr\"odinger equations with (space)-periodic potential)  that
\BEQ \left(\begin{array}{c} \psi_1(\theta+2\pi) \\ \psi_2(\theta+2\pi)\end{array}\right)
=M(u)\ .\ \left(\begin{array}{c} \psi_1(\theta)\\ \psi_2(\theta) \end{array}\right) \EEQ
for a certain matrix $M(u)\in SL(2,\R)$ (independent of $\theta$), called the 
{\it monodromy matrix}. Starting from  a different basis $\left(\begin{array}{c}
\tilde{\psi}_1\\ \tilde{\psi}_2 \end{array}\right)$, one obtains a conjugate matrix $\tilde{M}(u)$. The above action of the Virasoro group on the Hill equation leaves the monodromy matrix
unchanged, as can be seen from the transformed solutions $\phi_* \psi_1,\phi_* \psi_2$.
Hence the conjugacy class of the monodromy matrix is an invariant of the Hill operator
under the action of the diffeomorphism group.

Floquet's theory, together with the orbit theory for $SL(2,\R)$, imply that $\partial^2+u$ is
{\it stable} (meaning that all solutions are bounded) if $|\Tr M|<2$ or equivalently, if
$M$ is elliptic, i.e. conjugate to a rotation matrix; {\it unstable}  (meaning that all solutions are 
unbounded) if $|\Tr M|>2$ or equivalently, if
$M$ is hyperbolic, i.e. conjugate to a Lorentz shift $\left(\begin{array}{cc} e^{\lambda}&\\& e^{-\lambda}
\end{array}\right)$, $\lambda>0$. If $|Tr M|=2$, then  $M$ can be shown to
 be conjugate either to $\pm \Id$ or to the unipotent matrix $\pm \left(\begin{array}{cc} 1 & 2\pi \\ 0& 1\end{array}\right)$; in the
latter case, $\partial^2+u$ is {\em semi-stable}, with stable {\em and} unstable solutions.
Two linearly independent $2\pi$- or $4\pi$-periodic solutions exist when $M=\pm\Id$; only one in the unipotent case; and none
in in the remaining cases.

An important result due to Lazutkin-Pankratova (see \cite{LazPan75}) states that all stable Hill operators
are conjugate by a suitable time-reparametrization to a Hill operator with constant potential $\partial^2+\alpha$,
$\alpha>0$. They also distinguish between {\it oscillating} and {\it non-oscillating} equations (oscillating equations have
solutions with infinitely many zeros, while non-oscillating equations have solutions with at most one zero), but we shall
not need to go further into this. Let us just remark that (as they also show) non-oscillating operators are also conjugate to
a Hill operator with constant potential $\partial^2+\alpha$, with $\alpha\le 0$ this time. Hence operators of type II, resp.
III of Kirillov's classification (see Definition 2.3.4 below)
 are exactly the unstable, resp. semi-stable {\it oscillating} operators.

\bigskip

A {\it complete classification} of the orbits under the action of $\Diff_+(\R/2\pi/Z)$  may be obtained
by considering  the {\it lifted monodromy}. Set $\left(\begin{array}{c}
\psi_1(\theta) \\ \psi_2(\theta) \end{array}\right)=M(u)(\theta) \left(\begin{array}{c}
\psi_1(0)\\ \psi_2(0) \end{array}\right).$ The path $\theta\to M(u)(\theta)\in SL(2,\R)$
may be lifted uniquely to a path $\theta\to \tilde{M}(u)(\theta)\in \widetilde{SL}(2,\R)$ such that $M(u)(0)=\Id$, 
where $\widetilde{SL}(2,\R)$ is the universal covering of $SL(2,\R)$.  This procedure defines a unique  lifted
monodromy matrix $\tilde{M}(u):=\tilde{M}(u)(2\pi)$ modulo conjugacy.

 The following 
arguments (see \cite{KheWen}) show briefly why this invariant suffices to characterize the orbit of $u$
under diffeomorphisms. Set $\left(\begin{array}{c} \psi_1(\theta)\\ \psi_2(\theta)\end{array}\right)=
\sqrt{\xi(\theta)} \left(\begin{array}{c} \cos \omega(\theta) \\ \sin\omega(\theta) \end{array}\right)$. The Wronskian
$$W:=\psi_1\psi'_2-\psi'_1\psi_2$$
(a constant of motion) is equal to $\omega'(\theta)\xi(\theta) $, hence $\omega'=\frac{W}{\xi}$
is of constant sign, say $>0$ (by choosing $W>0$). By the action of $\Diff(\R/2\pi\Z)$,
one can arrange that $\omega'$ is constant, while $\omega(0)$ and $\omega(2\pi)$ remain
related by the homographic action of $M(u)$, viz. $\cotan \omega(2\pi)=\frac{a\cotan \omega(0)+b}{c\cotan \omega(0)+d}$ if $M(u)=\left(\begin{array}{cc} a &b\\c&d\end{array}\right)\in SL(2,\R)$. The lifting of the monodromy produces a supplementary invariant: the {\it 
winding number} $n:=\lfloor(\omega(2\pi)-\omega(0))/2\pi\rceil=\lfloor \frac{W}{2\pi} \int_0^{2\pi} \frac{d\theta}{\xi(\theta)} \rceil$
 ($\lfloor\ .\ \rceil$=entire part), namely, the integer number of complete rotations made by the angle $\omega$.

This change of function is particularly relevant in the {\it elliptic} case. Choose a basis
$\left(\begin{array}{c} \psi_1 \\ \psi_2 \end{array}\right)$ such that $M=\left(\begin{array}{cc}
\cos \lambda & -\sin\lambda \\ \sin\lambda & \cos\lambda \end{array}\right)$. Then
$\pm\lambda=\omega(2\pi)-\omega(0)=W\int_0^{2\pi} \frac{d\theta}{\xi(\theta)} \ [2\pi].$

If $M=\left(\begin{array}{cc} e^{\lambda} & \\ & e^{-\lambda} \end{array}\right) $ is {\it hyperbolic} instead, set rather 
\BEQ \psi_1^2(\theta)=\half |\xi(\theta)| e^{2\omega(\theta)},\quad \psi_2^2(\theta)=\half |\xi(\theta)| e^{-2\omega(\theta)}
\EEQ
with $\xi(\theta)=2(\psi_1\psi_2)(\theta)$, so that $\pm\lambda=\omega(2\pi)-\omega(0)\ [2\II\pi]$. Then one finds
$\omega'=-\frac{W}{\xi}$, hence $\omega=-W \int \frac{d\theta}{\xi(\theta)}$. The functions
$\frac{1}{\xi}$ and $\omega$ are not well-defined if $\psi_1$ or $\psi_2$ has some zeros. Supposing $u$ is analytic,
the functions $\psi_1,\psi_2$ may be extended analytically to some strip $\Omega=\{|\Im \theta|<\eps\}$.
Choose some contour $\Gamma\subset\Omega$ avoiding the zeros of $\psi_1$ and $\psi_2$ such that (assuming
$\xi(0)\not=0$, otherwise use a translation) $\Gamma(0)=0$ and $\Gamma(2\pi)=2\pi$. The idea is to keep $\Gamma$ {\em
real} away from some symmetric neighbourhood $U_{\eps}$ of the zeros, and to complete the path with
half-circles centered on the real axis of radius $\eps$ around each zero, taken indifferently in the upper- or
lower-half plane (compare with section 3.2 below where more care is needed). Suppose $\psi_1(\theta_0)=0$ for instance,
 so $\psi'_1(\theta_0)=a\not =0$ and $\psi_2(\theta_0)=-\frac{W}{a}$. Then
\BEQ -W \int_{\Gamma\cap [\theta_0-\eps,\theta_0+\eps]}  \frac{d\theta}{\xi(\theta)}=-W\int_{\theta_0-\eps}^{\theta_0+\eps}
 \frac{ d\theta}{\theta-\theta_0\pm \II 0} \frac{\theta-\theta_0}{\xi(\theta)}=-W\ p.v. \int_{\theta_0-\eps}^{\theta_0+\eps}
  \frac{d\theta}{\xi(\theta)} \pm \II \frac{\pi}{2} \EEQ
(depending on the position of the half-circle with respect to the real axis)
since $\frac{1}{\theta-\theta_0\pm\II 0}=p.v. \frac{1}{\theta-\theta_0} \mp \II \pi \del_{\theta_0}$ (see for instance \cite{GelShi}) 
 and the residues
of $\frac{1}{\xi(\theta)}$ at the zeros of $\xi$ are  $\pm \frac{1}{2W}$. It is clear from the above 
definitions that $\xi$ has only simple zeros, in {\em even} number.    Hence
$-W\int_{\Gamma} \frac{d\theta}{\xi(\theta)}\equiv -W \ p.v. \int_0^{2\pi} \frac{d\theta}{\xi(\theta)} \equiv
\lambda \ [\II\pi].$ By exponentiating, one obtains a
monodromy matrix in $PSL(2,\R)=SL(2,\R)/\{\pm 1\}$.

Finally, if $M$ is {\it unipotent}, $M=\pm \left(\begin{array}{cc} 1 & a \\ 0& 1\end{array}\right)$ in some
 basis $\left(\begin{array}{c} \psi_1 \\ \psi_2\end{array}\right)$, set $\psi_1(\theta)=\omega \psi_2(\theta)$
and $\xi=\psi_2^2$, so that $\omega(2\pi)=\omega(0)+a$. Then $\omega'=-\frac{W}{\xi}$, so $\omega$ is once again defined as 
 $-W\int \frac{d\theta}{\xi(\theta)}$ if $\xi$ does not have any zero. In the contrary case,
one uses a deformation of contour as in the hyperbolic case, to obtain
\BEQ -W\int_{\Gamma\cap[\theta_0-\eps,\theta_0+\eps]}  \frac{d\theta}{\xi(\theta)}=-W\int_{\theta_0-\eps}^{\theta_0+\eps}
 \frac{ d\theta}{(\theta-\theta_0\pm\II 0)^2} \frac{(\theta-\theta_0)^2}{\psi_2^2(\theta)}.\EEQ
Since $\frac{1}{(\theta-\theta_0\pm\II 0)^2}=p.v. \frac{1}{(\theta-\theta_0)^2} \pm \II \pi \del'_{\theta_0}$
and $\frac{(\theta-\theta_0)^2}{\psi_2^2(\theta)}=1+O((\theta-\theta_0)^2)$ -- since $\psi''_2(\theta_0)=-V_2(\theta_0)\psi_2(\theta_0)=0$ --,
 the Dirac term does not make
any contribution at all this time, hence 
\BEQ a=\omega(2\pi)-\omega(0)=-W \int_{\Gamma} \frac{d\theta}{\xi(\theta)},\EEQ
where $\Gamma:[0,2\pi]\to \C$ is an arbitrary contour as defined above. 

\bigskip

Summarizing:

{\bf Proposition 2.2.1}  (see \cite{KheWen} for (ii))

{\em
\begin{itemize}
\item[(i)]
The lifted monodromy of the operator $\partial^2+u$ is characterized by the (correctly normalized) quantity $\int_0^{2\pi}
 \frac{d\theta}{\xi(\theta)}$ or $\int_{\Gamma} \frac{d\theta}{\xi(\theta)}$, where $\xi\in \Stab_u$.
\item[(ii)]
The orbits under the diffeomorphism group of the space of Hill operators are characterized
by the conjugacy class of their lifted monodromy. More precisely, the lifted monodromy
defines a bijection from the set of orbits onto the space of conjugacy classes of
$\left(\widetilde{SL}(2,\R)\setminus\{\pm1\}\right)/\{\pm 1\}$ (an element $M\in \widetilde{SL}(2,\R)$ has to 
be identified with its opposite $-M$).
\end{itemize}
}

\subsection{Kirillov's classification of Hill operators by isotropy subgroups}

Another classification, also useful for our purposes (and more
explicit in some sense), is due to Kirillov. Introduce first

{\bf Definition 2.3.1}
{\em

Let  $Stab_u$, $u\in C^{\infty}(\R/2\pi\Z)$  be the {\em isotropy subgroup} (or {\em stabilizer}) of $\partial^2+u$ in $\Diff_+(\R/2\pi\Z)$, namely,
\BEQ Stab_u:=\{ \phi\in \Diff_+(\R/2\pi\Z)\ |\ \phi_* (\partial^2+u)=\partial^2+u\}.\EEQ
}

{\bf Proposition 2.3.2 (definition of the first integral $I$)}  (see \cite{Gui94})

{\em
\begin{enumerate}
\item
Let $\xi\in C^{\infty}(\R/2\pi\Z)$: then $\xi\in Lie(Stab_u)$ if and only if $\xi$
satisfies
\BEQ \half \xi'''+2u\xi'+u'\xi=0.\EEQ
\item
Let $I_u(\xi):=\xi \xi''-\half \xi'^2+2u\xi^2$. Then $I_u(\xi)$ is a constant of 
motion if $\xi\in Lie(Stab_u)$.
\item
Consider $\phi\in\Diff_+(\R/2\pi\Z)$ and the transformed potential $\tilde{u}$ such
that $\phi_*(\partial^2+u)=\partial^2+\tilde{u}$. Then
\BEQ I_{\tilde{u}}(\phi'^{-1}\ .\  \xi\circ\phi)=I_u(\xi).\EEQ
\item
Consider the Hill equation $(\partial^2+u)\psi(\theta)=0.$ If $(\psi_1,\psi_2)$ is a basis
of
solutions of this equation, then $\xi:=a_{11} \psi_1^2+2 a_{12} \psi_1\psi_2+a_{22} \psi_2^2$  ($a_{11},a_{12},a_{22}\in\R$)
 satisfies the equation
\BEQ \half \xi'''+2u\xi'+u'\xi=0 \label{iso3} \EEQ
In other terms, $\xi\in Stab_u$ is in the isotropy subgroup  of the Hill operator $\partial^2+u$.

Conversely, any solution  of (\ref{iso3}) can be obtained in this way.
\item (same notations) consider in particular $\xi=\psi_1^2+\psi_2^2$. Then $I_u(\xi)=W^2$ if $W$ is
the Wronskian of $(\psi_1,\psi_2)$, namely, $W=\psi_1 \psi'_2-\psi'_1\psi_2$ (constant of the motion).

\end{enumerate}
}

Note (see 3.) that $\left(\phi'^{-1}\ .\  \xi\circ\phi\right)\partial$
 is the conjugate of the vector $\xi\partial\in Vect(S^1)$ by the diffeomorphism $\phi$. Hence one may say that the first integral $I$ is invariant
under the (adjoint-and-coadjoint) action of $\Diff_+(\R/2\pi\Z)$.

Consider now the (adjoint) orbit of $\xi$ under $\Diff_+(\R/2\pi\Z)$. Clearly,
$\int_0^{2\pi} \frac{d\theta}{\xi(\theta)}$ (if well-defined, i.e. if $\xi$ has no zero) does not
depend on the choice of the point on the orbit since $\int_0^{2\pi} \frac{d\theta}{\phi'^{-1}(\theta) \xi\circ\phi(\theta)} =\int_0^{2\pi} \frac{du}{\xi(u)}.$ It is easy to see from Prop. 2.3.2(2)
that $\xi$ either never vanishes ({\it case I}), or has an even number of simple zeros
({\it case II}), or has a finite number of double zeros ({\it case III}). Cases II, III correspond to a hyperbolic,
resp. unipotent monodromy matrix (see discussion in section 2.2).  In case II,
$I_u(\xi)=-\half \xi'(t_0)^2<0$ if $t_0$ is any zero. The principal value integral
$p.v. \int_0^{2\pi} \frac{dt}{\xi(t)}$ is well-defined. In case III, $I_u(\xi)=0$ and
the regularized integral $\int_{\Gamma} \frac{d\theta}{\xi(\theta)} $
(see above) is well-defined and independent of the choice of the contour $\Gamma$. Note that A. Kirillov uses instead the
following regularization,
$\lim_{\eps\to 0} \int_{[0,2\pi]\setminus U_{\eps}} \frac{dt}{\xi(t)}-\frac{C}{\eps}$ (where $U_{\eps}$ is a symmetric $\eps$-neighbourhood of the zeros) with $C$ chosen so that the limit is finite. The two regularizations are different. Both
are perfectly satisfactory to define an invariant of the orbits, but computations show that the Berry phase
is proportional to $\int \frac{d\theta}{\xi(\theta)}$.

Now the integral $\int_0^{2\pi} \frac{d\theta}{\xi(\theta)}$ (case I) and its variants for case II, III
are invariants under the diffeomorphism group. The discussion in section 2.2  shows that they characterize the lifted monodromy
of $\partial^2+u$. The
invariant $I_u(\xi)$ is also needed to fix $u$ uniquely in case I (see Prop. 2.3.2(2)) since $\xi$ stabilizes all
operators of the type $\partial^2+u+\frac{C}{\xi^2}$ ($C\in\R$). It turns out that $\int_0^{2\pi}$ -- or its variants -- and
$I_u(\xi)$ (in cases II and III), together with a discrete invariant $n\in\N$, suffice to distinguish between the 
different adjoint orbits of stabilizers (note that general adjoint orbits may be much more complicated, see \cite{GuiRog}). One has
the following:

{\bf Proposition 2.3.3 - Classification of the coadjoint invariants and of the orbits}  (see Kirillov \cite{Kir82})

{\em
\begin{enumerate}
 
\item Case I: $\xi$ is conjugate by a diffeomorphism $\phi$ to a (non-zero) constant $a\partial_{\theta}$, $a\not=0$.
 Hence $\phi'^{-1} \ .\  \xi\circ\phi\in
Lie(Stab_{\partial^2 + \alpha})$ for a certain constant $\alpha$. The stabilizer $Stab_{\partial^2+\alpha}$
is:

\begin{itemize}
\item[(i)] (non-generic case) either isomorphic to $\widetilde{SL}^{(n)}(2,\R)$ (the $n$-fold covering of $SL(2,\R)$),
with $Lie(Stab_{\partial^2+\alpha})=\R \partial_{\theta} \oplus \R \cos n\theta\partial_{\theta} \oplus \R
\sin  n\theta \partial_{\theta}$ if $\alpha=\frac{n^2}{4}$ for some $n\in\N^*$; then the monodromy in $PSL(2,\R)=SL(2,\R)/\{\pm 1\}$ is 
trivial, while the lifted monodromy matrix is the central element in $\widetilde{SL}(2,\R)/\{\pm 1\}$ corresponding
to a rotation of an angle $\pi n$;
\item[(ii)] or (generic case)  one-dimensional, equal to the rotation group $Rot\subset \Diff_+(\R/2\pi\Z)$
generated by the constant field $\partial$ in the remaining cases.

The  invariants are given by  $I_u(\xi)=2\alpha a^2$, $\int_0^{2\pi} \frac{d\theta}{\xi(\theta)}=\frac{2\pi}{a}$. The 
monodromy can be in any conjugacy class of $PSL(2,\R)$ except $\pm \Id$.
\end{itemize}

\item Case II: $\xi$ is conjugate to the field $a\sin n\theta(1+\alpha \sin n\theta)\partial_{\theta}$, $n=1,2\ldots$, 
$0\le \alpha<1$, which stabilizes $\partial^2+u_{n,\alpha}$, where
\BEQ u_{n,\alpha}(\theta):= \frac{n^2}{4} \left[\frac{1+6\alpha \sin n\theta+4\alpha^2 \sin^2 n\theta}{(1+\alpha\sin n\theta)^2}\right].\EEQ
The monodromy matrix is hyperbolic.
The invariants take the values $I_u(\xi)=-2a^2 n^2<0$, p.v.$\int_0^{2\pi} \frac{d\theta}{\xi(\theta)}=\frac{2\pi\alpha}{a\sqrt{1-\alpha^2}}.$
\item Case III: $\xi$ is conjugate to $\xi_{\pm,n,\alpha}:=\pm(1+\sin n\theta)(1+\alpha \sin n\theta)\partial$,
$0\le \alpha<1$, corresponding to a potential $v_{n,\alpha}$,
\BEQ v_{n,\alpha}(\theta)=\frac{n^2}{4} \left[ \frac{(\alpha-1)^2 +2\alpha(3-\alpha)\sin n\theta+4\alpha^2 \sin^2 n\theta}{(1+\alpha \sin n\theta)^2} \right] \label{vnalpha} \EEQ
The monodromy matrix is unipotent. The invariant $I_u(\xi)$ vanishes, while $\int_{\Gamma} \frac{d\theta}{\xi_{+}(\theta)}=
\frac{-2\pi}{(1-\alpha)\sqrt{1-\alpha^2}}.$ The discrete invariant $n$ suffices to characterize the orbit of $\partial^2+u$.
\end{enumerate}
In cases II and III (provided $\alpha>0$), the stabilizer is one-dimensional, generated by $\xi\partial_{\theta}$.

In the generic cases (case I, $\alpha\not= n^2/4$, $n=0,1,\ldots$ or case II) the monodromy matrix is elliptic, resp. hyperbolic, if and only if $I_u(\xi)>0$, resp. $I_u(\xi)<0$. In cases I ($\alpha=0$) and  III (with unipotent monodromy), $I_u(\xi)=0$.
}

There is a mistake in Lemma 3 of \cite{Kir82} (the potential $u_{n,\alpha}$ given
there is not correct). The potential $v_{n,\alpha}$ was missing, together with the value of $\int_{\Gamma} \frac{d\theta}{\xi_{\pm}(\theta)}.$
Both are obtained by straightforward computations.

This classification is also natural when one thinks of the behaviour of the solutions (see Lazutkin-Pankratova \cite{LazPan75}
and section 2.2).
In particular, case II (resp. III) correspond to operators with unstable (resp. semi-stable), oscillating solutions, while 
case I corresponds to operators with stable, oscillating solutions ($\alpha>0$), resp. unstable, non-oscillating solutions
($\alpha<0$), resp. semi-stable, non-oscillating solutions $(\alpha=0)$.

Note that in the case I generic, the three-dimensional isotropy subalgebra contains fields $\xi$ of 
type I, II $(\alpha=0)$ and III $(\alpha=0)$, hence the following nomenclature:

{\bf Definition 2.3.4}
{\em

If $\partial^2+u$ has a stabilizer $\xi$ of type I, or of type II, III with $\alpha=0$, then $\partial^2+u$
may be turned into a Hill operator with constant potential, and we shall say that the operator
$\partial^2+u$ (or the potential $u$) is of type I. If $\partial^2+u$ has a stabilizer
of type II, resp. III with $\alpha\not=0$, then we shall say that $\partial^2+u$ and $u$ are of type II, resp.
type III.

Similarly, we shall say that the Schr\"odinger operator $-2\II \partial_{\theta}-\partial_x^2+V_2(\theta) x^2+V_1(\theta)  x+V_0(\theta)$
is of type I (resp. II, III) if the Hill operator $\partial_{\theta}^2+V_2(\theta)$ is of the corresponding type.
}

Note that the cases I generic ($\alpha\not=\frac{n^2}{4},n=0,1,\ldots$) and II are generic (i.e. dense in ${\cal S}_{\le 2}^{aff}$).

\vskip 1 cm

Now the eigenvalues of the monodromy matrix (and also the lifted monodromy) can easily be obtained once one knows the values of the invariants
$\int_0^{2\pi} \frac{d\theta}{\xi(\theta)}$ and $I_u(\xi)$. The following Lemma gives the link between the two classifications:

{\bf Lemma 2.3.5}

{\em
Suppose $D=\partial^2+u$ is of type I (with $\alpha\not=0$) or II (i.e. its monodromy is either elliptic or hyperbolic).
If $D$ is of type I non generic, conjugate to $\partial^2+n^2/4$ for some $n\ge 1$,  choose $\xi$ to be conjugate to 
some non-zero multiple of $\partial_{\theta}$. Now (in all cases) normalize $\xi$ by requiring that $I_u(\xi)=2$, so that
$\xi$ is real in the elliptic case and purely imaginary in the hyperbolic case. Then the eigenvalues of the monodromy matrix are given by
$\exp \pm\II \int_0^{2\pi} \frac{d\theta}{\xi(\theta)}$ or
$\exp \pm\II \ p.v.\int_0^{2\pi} \frac{d\theta}{\xi(\theta)}$.
}

{\bf Proof.}

Coming back to the discussion in section 2.2, one  checks easily (with the normalization chosen there) that $I_u(\xi)=2W^2$ in the elliptic case,
and $I_u(\xi)=-2W^2$ in the hyperbolic case. Choose a basis of solutions $(\psi_1,\psi_2)$ such that $W=1$ and multiply
$\xi$ by $\II$ in the hyperbolic case. Then (in both cases) the eigenvalues of the monodromy matrix ($\pm \II \lambda$ in the
elliptic case, and $\pm \lambda$ in the hyperbolic case) are given by $\exp \pm \II  \int_0^{2\pi} \frac{d\theta}{\xi(\theta)}$ or the exponential of the corresponding principal value integral.
\hfill \eop

\subsection{Classification of the SV-orbits in ${\cal S}_{\le 2}^{aff}$}

This problem can be solved by extending the above results, which may be interpreted as the decomposition
of ${\cal S}_2^{aff}$ into $\Diff_+(\R/2\pi\Z)$-orbits.
 Let us first compute the stabilizers of some operators that will be shown later to  be representatives of all the orbits.
We choose to present the results in the Fourier coordinates $(\theta,x)$. The orbits of type I, resp. III split into orbits of
type (i), (i)bis, resp. (iii), (iii)bis due to the presence of the linear term $V_1(\theta)x$ in the potential.

The computations depend on the formulas of Proposition 1.5, see formulas (\ref{actphi}), (\ref{actab}),
(\ref{actdphi}), (\ref{actdab}) for more convenience.

{\bf Definition 2.4.1}
{\em

If $D\in {\cal S}_{\le 2}^{aff}$, we denote by $G_D$ the stabilizer of $D$ in the Schr\"odinger-Virasoro
group SV, i.e. $G_D=\{ g\in \SV\ |\ \sigma_{1/4}(g).D=D\}.$
}

Recall the notation $\Stab_u$, $u\in C^{\infty}(\R/2\pi\Z)$ is used for the stabilizer in $\Diff_+(\R/2\pi\Z)$ of the corresponding Hill operator.

 Note  that ${\cal M}_1=M_0$ (whose exponential amounts to the multiplication of
 the wave functions  $\psi$  by
a constant phase) acts trivially on any operator $D$, hence ${\cal M}_1\in G_D$ always.
The rotation group $\theta\to \theta+\theta_0$ generated by $d\sigma_{1/4}({\cal L}_1)=d\sigma_{1/4}(L_0)=-
\partial_{\theta}$ will be denoted by $\Rot$.

In the following classification, we shall call {\em harmonic oscillators} (resp. {\em harmonic repulsors}) operators with elliptic,
resp. hyperbolic monodromy.

\begin{itemize}
\item[(i)] \underline{ Time-independent harmonic oscillators or repulsors}

Set $D_{\alpha,\gamma}:=-2\II \partial_{\theta}-\partial_x^2+\alpha x^2+\gamma$
($\alpha,\gamma\in\R$). It is clear that $L_{-1}=\partial_{\theta}$ leaves $D_{\alpha,\gamma}$ invariant in all cases. Suppose first for simplicity that $\gamma=0$. Then $G_D=(G_0)_D \ltimes H_D$ (see Proposition 1.3 for notations) is a semi-direct
product, so one retrieves Kirillov's results (see Proposition 2.3.3,  case I) for $(G_0)_D$; to be specific,
$Lie((G_0)_{D_{n^2/4,0}})=\R\partial_{\theta}\oplus \R {\cal L}_{\sin n\theta}\oplus
\R {\cal L}_{\cos n\theta}$ if $n\in\N^*$, and $Lie((G_0)_{D_{\alpha,0}})=\R\partial_{\theta}$ otherwise.

Now $(1;(a,b))\in H_{D_{\alpha,0}}$ if and only if $b'=0$ and $a''=-\alpha a$. The latter equation
has a non-trivial solution if and only if $\alpha=0$ (in which case
$Lie (H_D)=\R {\cal Y}_1\oplus \R {\cal M}_1$) or  $\alpha=n^2/4$, $n\ge 1$ with $n$ {\it even},
in which case $Lie(H_D)=\R {\cal Y}_{\cos n\theta/2} \oplus \R {\cal Y}_{\sin n\theta/2} 
\oplus {\cal M}_1$. Then $\exp \frac{1}{n} {\cal L}_1\subset \widetilde{SL}^{(n)}(2,\R)$ is the rotation of angle $2\pi$, while 
$\exp \frac{1}{n} \ad {\cal L}_1|_{[Lie({\cal H}_D), Lie({\cal H}_D)]}$ is a rotation of angle $\pi$.

The isotropy groups $G_D$ are the same in the case $\gamma\not=0$, except for a different embedding involving
sometimes complicated components in the nilpotent part of SV which do not change the commutation relations (so that
$G_D$ is no more a semi-direct product $(G_0)_D\ltimes H_D$).

All together, one has proved:

{\bf Theorem 2.4.2}

{\em
\begin{enumerate}

\item  If $\alpha=n^2/4$, where $n\ge 2$ is an even integer, then
$G_{D_{n^2/4,\gamma}}\simeq \widetilde{SL}^{(n)}(2,\R)\ltimes {\cal H}_1$ is isomorphic to an $n$-covering
of the Schr\"odinger group; the semi-direct action of  
$\widetilde{SL}^{(n)}(2,\R)$ quotients out into an action of the two-fold covering  $\widetilde{SL}^{(2)}(2,\R)$.
  The Lie algebra of
the group $\widetilde{SL}^{(n)}(2,\R)$ acts as $\R \partial_{\theta}\oplus \R ({\cal L}_{\sin n\theta}+
 {\cal M}_{-\half \gamma \sin n\theta}) \oplus \R ( {\cal L}_{\cos n\theta}+  {\cal M}_{-\half
\gamma\cos n\theta})$.   After transformation to the
Laurent coordinates $(t,r)$ (and supposing $\gamma=0$), $G_{D_{n^2/4,0}}$ is the connected Lie group with Lie algebra
$\langle L_0,L_{\pm n} \rangle\ltimes \langle Y_{\pm n/2},M_0\rangle\subset \sv(0)$.

\item  If $\alpha=n^2/4$, where  $n\ge 1$
is odd, then
$G_{D_{n^2/4,0}}\simeq \widetilde{SL}^{(n)}(2,\R)\times \exp \R {\cal M}_1$.

\item
If $\alpha=0$,  then
$G_{D_{0,\gamma}}=Rot\times \exp(\R {\cal Y}_1\oplus \R {\cal M}_1) \simeq (\R/2\pi\Z)\times \R\times (\R/2\pi\Z)$ is the commutative group of  constant translations-phases. After transformation to the Laurent
coordinates $(t,r)$, it is the connected Lie group with Lie algebra $\langle L_0, Y_0, M_0\rangle\subset\sv(0)$.

\item 
In the generic case $\alpha\not=n^2/4$, $n=0,1,\ldots$ one has simply $G_D=\Rot\times\exp\R {\cal M}_1\simeq (\R/2\pi\Z)^2$.

\end{enumerate}
}

It is natural in view of these results to consider the two-fold covering
$\tilde{H}^{(2)}$ of $H$ obtained by considering $4\pi$-periodic fields. Then the stabilizer
in $\widetilde{SV}^{(2)}:=G_0\ltimes \tilde{H}^{(2)}$ of $D_{n^2/4,0}$ ($n\ge 1$ odd)
is isomorphic to $\widetilde{SL}^{(n)}(2,\R)\ltimes {\cal H}_1$ as in the case of an 
even index $n$. This time $Lie({\cal H}_1)= \langle Y_{\pm n/2},M_0\rangle\subset \sv(\half)$.

The  best-known case is $\alpha=1/4$ $(n=1)$, $\gamma=0$. In the Laurent coordinates
$(t,r)$, $ D_{1/4,0}$ writes $-2\II \partial_t-\partial_r^2$, namely, it is the
free Schr\"odinger equation. Then $SL(2,\R)\ltimes {\cal H}_1$ acts on $D_{1/4,0}$ in the usual way (see formulas
 (\ref{repsv0})) in the Laurent coordinates.

\item[(i)bis]  \underline{Special time-independent harmonic oscillators with added resonant oscillating drift}

Consider 
$$D=-2\II \partial_{\theta}-\partial_x^2+n^2 x^2+C\cos(n\theta-\sigma/2)\ .\ x+\gamma$$ ($C,\sigma,\gamma\in\R$,
 $C\not=0$, $n\ge 1$ integer).
Then computations show that $G_D\simeq \R\times\R\times \R/2\pi\Z$ is three-dimensional, generated by
\BEQ {\cal L}_{1-\cos(2n\theta-\sigma)} + {\cal Y}_{\frac{C}{8n} \sin 3(n\theta-\sigma/2)} -
{\cal M}_{\frac{C^2}{32n^2}(\frac{\cos 4n(\theta-\sigma/2)}{4}+\frac{\cos(2n\theta-\sigma)}{2}
+\frac{\gamma}{2} \cos(2n\theta-\sigma))} ,\EEQ
\BEQ {\cal Y}_{C\sin(n\theta-\sigma/2)}+{\cal M}_{\frac{C^2}{8n}\cos(2n\theta-\sigma)} \EEQ
and ${\cal M}_1$. One checks (by direct computation) that the value of the associated invariant $I_{n^2}(1-\cos(2n\theta-\sigma))$ is $0$.

\item[(ii)] \underline{Time-dependent Ince harmonic repulsors of type II}

Consider 
\BEQ D_{n,\alpha,\gamma}=-2\II\partial_{\theta}-\partial_x^2+u_{n,\alpha}(\theta) x^2+\gamma,\quad n=1,2,\ldots,\  \alpha\in(0,1) \EEQ
where
\BEQ u_{n,\alpha}(\theta)=  \frac{n^2}{4} \left[\frac{1+6\alpha \sin n\theta+4\alpha^2 \sin^2 n\theta}{(1+\alpha\sin n\theta)^2}\right].\EEQ

Then (see preceding subsection) $ {\cal L}_{\xi} -\frac{\gamma}{2} {\cal M}_{\xi} \in
Lie(G_D)$  $(\xi\not=0)$ if and only if  $\xi$ is proportional to $\xi_{n,\alpha}$, with $\xi_{n,\alpha}=\sin n\theta(1+\alpha\sin n\theta)\partial_{\theta}$.  Now
$$d\sigma_{1/4}({\cal Y}_{f_1}+{\cal M}_{f_2}).D=0$$
if and only if $f'_2=0$ and $f_1''+u_{n,\alpha}f_1=0$. The latter equation is known under
the name of {\it Ince's equation} (see Magnus and Winkler, \cite{MagWin}). The change of  variable and function
$\theta\to \del(\theta)=\frac{\pi}{4}-n\frac{\theta}{2}$, $f_1(\theta)\to y(\del)=(1+\alpha\cos 2\del)^{b/4\alpha} f_1(\del(\theta))$ with $b=-2\alpha[1+\frac{\II \alpha}{\sqrt{1-\alpha^2}}]$  turns the above equation into the standard form
$$(1+a\cos 2\del) y''+b\sin 2\del y'+(c+d\cos 2\del)y=0$$ with $a=\alpha$, $c=1-\frac{\alpha^2}{1-\alpha^2}$,
$d=\alpha[3+\frac{\alpha^2}{1-\alpha^2}\mp \frac{2\II\alpha}{\sqrt{1-\alpha^2}}]$.
 Conditions for the coexistence
of two independent periodic solutions of Ince's equation have been studied in detail. In our case, there is
no periodic solution since $\partial^2+u_{n,\alpha}$ is  unstable (see discussion in section 2.2).  Hence
\BEQ G_D=\exp ( \R ( {\cal L}_{\xi_{n,\alpha}}-\frac{\gamma}{2} {\cal M}_{\xi_{n,\alpha}})  \oplus\R {\cal M}_1)\simeq \R\times (\R/2\pi\Z). \EEQ

\item[(iii)] \underline{Non-resonant time-dependent Schr\"odinger operators of type III}

Consider $$D_{n,\alpha,\gamma}=2\II \partial_{\theta}-\partial_x^2+v_{n,\alpha}(\theta)x^2+\gamma$$
$(n=1,2,\ldots, \alpha\in(0,1))$. See equation (\ref{vnalpha}). 
Similarly to case (ii), $ {\cal L}_{\xi} -\frac{\gamma}{2} {\cal M}_{\xi} \in
Lie(G_D)$  $(\xi\not=0)$ if and only if  $\xi$ is proportional to $\xi_{\pm,n,\alpha}$, where
 $\xi_{\pm,n,\alpha}=\pm(1+\sin n\theta)(1+\alpha\sin n\theta)\partial_{\theta} $.  Then $d\sigma_{1/4}({\cal Y}_{f_1}+{\cal M}_{f_2}).D=0$
if and only if $f'_2=0$ and $f_1''+v_{n,\alpha}f_1=0$. This is once again Ince's equation, with parameters
$a=\alpha$, $b=-2\alpha$, $c=1-2\alpha$, $d=3\alpha$. One verifies immediately that $y(\del)=\cos \del$ is
the unique (up to a constant)  periodic solution of this semi-stable Hill equation, corresponding to  $f_1(\theta):=(1+\alpha\sin n\theta)^{\half} \cos(\frac{\pi}{4}-n\frac{\theta}{2})$. Note that $\xi_{+,n,\alpha}=f_1^2$ (so that $f_1$ is - up to a sign - the unique
$C^{\infty}$ square-root of $\xi_{+,n,\alpha}$). Hence
\BEQ G_D=\exp\left(\R ({\cal L}_{\xi_{\pm,n,\alpha}}-\frac{\gamma}{2} {\cal M}_{\xi_{\pm,n,\alpha}}) \oplus \R {\cal Y}_{f_1}\oplus\R{\cal M}_1\right)\simeq
\R\times\R\times(\R/2\pi\Z) \EEQ 

\item[(iii)bis] \underline{Schr\"odinger operators of type III with added resonant drift}

Consider 
$$D=-2\II\partial_{\theta}-\partial_x^2+v_{n,\alpha}(\theta)x^2+C(1+\alpha\sin n\theta)^{1/2} \cos(\frac{\pi}{4}-n\frac{\theta}{2}) x+\gamma$$
$(C\not=0)$ with $v_{n,\alpha}$ as in case (iii). Set $\xi(\theta)=(1+\sin n\theta)(1+\alpha \sin n\theta)$ and $f(\theta)=(1+\alpha\sin n\theta)^{\half} \cos(\pi/4-n\theta/2)$. 
Recall $\xi=f^2$.

 Suppose
${\cal L}_{\xi}+{\cal Y}_{f_1}+{\cal M}_{f_2}$ stabilizes $D$. Then (see (\ref{actdphi}))
\BEQ
    2(f''_1+v_{n,\alpha}f_1)=C(\xi f'+\frac{3}{2} \xi' f)=4C f^2 f' \label{eq(v)1} \EEQ
and
\BEQ f'_2=-\half(\gamma \xi'+ Cf_1 f) \label{eq(v)2} .\EEQ
The kernel of the operator $\partial^2+v_{n,\alpha}$ is one-dimensional, generated by $f$. Hence the above
equation (\ref{eq(v)1}) has a solution if and only if
$\int_0^{2\pi} (\xi f'+\frac{3}{2}\xi' f)f\ d\theta=0$, which is true since $(\xi f'+\frac{3}{2}\xi' f)f=(f^4)'$.
Now equation (\ref{eq(v)2}) has a solution if and only if $f_1$ is chosen to be the unique solution orthogonal
to the kernel of $\partial^2+v_{n,\alpha}$, namely, if $\int_0^{2\pi} f_1 f \ d\theta=0$.

Now
\BEQ d\sigma_{1/4}({\cal Y}_{g_1}+{\cal M}_{g_2}).D=-2(g''_1+v_{n,\alpha}g_1)x-fg_1-2g'_2 \EEQ
vanishes if and only if $g_1=f$ (up to a multiplicative constant) and $\int_0^{2\pi} g_1 f \ d\theta=0$. The two conditions
are clearly incompatible.

All together one has proved that $G_D=\exp \left(  \R ({\cal L}_{(1+\sin n\theta)(1+\alpha \sin n\theta)}+{\cal Y}_{f_1}
+{\cal M}_{f_2}) \oplus \R {\cal M}_1 \right) \simeq \R\times \R/2\pi\Z$ (with $f_1,f_2$ solving equations (\ref{eq(v)1}), (\ref{eq(v)2}))
 is commutative two-dimensional. 

Explicit but cumbersome formulas for $f_1,f_2$ are easy to derive from the proof of Lemma 4.10 below. We shall not need them.
\end{itemize}

There remains to prove that we have classified {\it all} the  orbits in ${\cal S}_{\le 2}^{aff}$.

{\bf Theorem  2.4.3}

{\em
Any Schr\"odinger operator $D$ in ${\cal S}_{\le 2}^{aff}$ belongs to the orbit of one of the above operators.
}

{\bf Proof.}

Let $D\in {\cal S}_{\le 2}^{aff}$. Suppose first that $V_2$ is of type I. Then one may assume (by a time-reparametrization) 
that $V_2=\alpha$ is a constant. The operator $D$ belongs to the orbit of $D_{\alpha,\gamma}$ (case (i)) for some $\gamma$ if 
and only if $V_1=2(a''+\alpha a)$. If $\alpha\not=n^2$ (or $n^2/4$ if one considers the $\widetilde{SV}^{(2)}$-orbits) then this
equation has a unique solution for every $V_1$. If $\alpha=n^2$, then a Fourier series $V_1=\sum_k c_k e^{\II k\theta}$ is
in the image of $\partial_{\theta}^2+\alpha$ if and only if $c_{\pm n}=0$. This analysis accounts for
the two cases (i), (i)bis. 

Suppose now $V_2$ is of type II. By a time-reparametrization one may choose $V_2=u_{n,\alpha}$. The operator $D$ belongs
to the orbit of $D_{n,\alpha,\gamma}$ (see case (ii)) for some $\gamma$, provided $V_1=2(a''+u_{n,\alpha}a)$. Since $\partial^2+u_{n,\alpha}$ 
(acting on $C^{\infty}(\R/2\pi\Z)$) has
a trivial kernel, it has a bounded inverse and the unique solution of the above equation is $C^{\infty}$. Hence $D$ belongs
to the orbit of $D_{n,\alpha,\gamma}$.

Finally, suppose $V_2$ is of type III. One is led to solve the equation $V_1=2(a''+v_{n,\alpha}a)$. Recall $V_1(\theta)=(1+\alpha \sin n\theta)^{1/2} \cos(\frac{\pi}{4}-n\frac{\theta}{2})$ solves the equation $f''_1+v_{n,\alpha}f_1=0$. Hence $V_1=2(a''+v_{n,\alpha}a)$
has a solution if and only if $\int_0^{2\pi} V_1(\theta) (1+\alpha \sin n\theta)^{1/2} \cos(\frac{\pi}{4}-n\frac{\theta}{2}) \ d\theta=0$, which accounts for cases (iii), (iii)bis. \hfill \eop

Note that Schr\"odinger operators of type III are generically of type (iii)bis, and Schr\"odinger operators of type I with
$\alpha=n^2$, $n=1,2,\ldots$ are generically of type (i)bis.

{\bf Corollary  2.4.4}

{\em
For generic orbits (type (i) with $\alpha\not=\frac{n^2}{4}$, $n\ge 0$, or type (ii)), the isotropy group is two-dimensional, given by $\exp \R({\cal L}_{\xi}+{\cal Y}_{\del_1}+{\cal M}_{\del_2})\oplus \R{\cal M}_1\simeq \R\times\R/2\pi\Z$ or $\R/2\pi\Z\times\R/2\pi\Z$ for some triple $(\xi,\del_1,\del_2)$ with $\xi\not=0$.
}

Let us finish with a remark. Consider a potential $V_2(\theta)x^2+V_1(\theta)x+V_0(\theta)$ 
of type (i), (ii) or (iii). As we shall see in the next section,
the monodromy of the corresponding Schr\"odinger operator  depends only on  the (conjugacy class of the)
invariant $\xi$ and the value of the constant $\gamma$ (which acts as a simple energy shift).
 Computing the invariant $\xi$ is a difficult task in general,
but suppose it can be achieved. How does one determine the constant $\gamma$ ? We give an  answer for generic elliptic
or hyperbolic potentials
of type (i).

{\bf Lemma 2.4.5}

{\em
Let $D=-2\II \partial_{\theta}-\partial_x^2+V_2(\theta)x^2+V_1(\theta)x+V_0(\theta)$ be of type (i), elliptic or
hyperbolic, generic, so that $D$ is conjugate to a unique operator $D_{\alpha,\gamma}=-2\II \partial_{\theta}-\partial_x^2
+\alpha x^2+\gamma$ ($\alpha\in\R,\ \alpha\not=\frac{n^2}{4},\ n=0,1,\ldots$). Then $\gamma$ may be retrieved from
\BEQ \gamma=\frac{1}{2\pi} \int_0^{2\pi} (V_0-\frac{1}{4} V_1 W_1)(\theta)\ d\theta \EEQ
where $W_1$ is the unique solution of the equation $(\partial^2+V_2)W_1=V_1$.
}

{\bf Proof.}

Start from the model operator $D_{\alpha,\gamma}$, with stabilizer $\xi=1$, and apply successively
$\sigma_{1/4}(\phi;(0,0))$ and $\sigma_{1/4}(1;(g,h))$. Then one obtains the operator
$$D:=-2\II\partial_{\theta}-\partial_x^2+V_2(\theta)x^2-2\left( (\partial^2+V_2)g\right)  x+(\gamma\dot{\phi}+g(\partial^2+V_2)g
-2\dot{h}) $$
(see formulas (\ref{actphi}),(\ref{actab})). Now $\int_0^{2\pi} \dot{\phi}(\theta)\ d\theta=2\pi$ since
$\phi\in \Diff_+(\R/2\pi\Z)$. Hence the result. \hfill \eop

\subsection{Connection to U. Niederer's results}

We are refering to a classical paper by U. Niederer (see \cite{Nie74}) concerning the
maximal groups of Lie symmetries of Schr\"odinger equations with arbitrary potentials.
One may rephrase his main result as follows (though the Schr\"odinger-Virasoro had not been introduced
at that time). U. Niederer shows that any transformation 
$$\psi(t,r)\to \tilde{\psi}(t,r)=\exp \II f_g(g^{-1}(t,r)) 
\psi(g^{-1}(t,r))$$ 
(where   $g:(t,r)\to (t',r')$
is an arbitrary coordinate transformation  and $f_g$ an arbitrary 'companion function'
corresponding to a projective action) carrying the space of solutions of the Schr\"odinger equation
\BEQ (-2\II \partial_t-\partial_r^2+V(t,r))\psi(t,r)=0 \label{NieSch} \EEQ
into itself in necessarily of the form
$\sigma_{1/4}(g)$ for some $g\in\ SV$. This is - by the way - an elegant way of introducing the SV group
in the first place. Then Niederer  gives a necessary and sufficient condition for $g\in \sigma_{1/4}(SV)$ to leave (\ref{NieSch}) invariant, and produces some physically interesting examples.
Let us analyze some of these examples from our point of view. It should be understood that 
Niederer's examples are given in the Laurent coordinates $(t,r)$ and should hence be transformed by using
Lemma 1.6 to compare with our results.

\begin{itemize}
\item[(i)] $V=0$ (free Schr\"odinger equation): this case corresponds after the transformation at the end of Section 1
to the potential $V(\theta,x)=\frac{1}{4} x^2$, with invariance under the full Schr\"odinger
group (see case (i) in subsection 2.4, with $\alpha=1/4$ and $\gamma=0$).
\item[(ii)] $V=-gr$ (free fall) corresponds to $V(\theta,x)=\frac{1}{4} x^2-ge^{\II
(\theta/2+3\pi/4)} x$ (a $4\pi$-periodic potential), which belongs to the same orbit as case
(i) (free Schr\"odinger equation in the Laurent coordinates). 
\item[(iii)] $V=\half \omega^2 r^2$ (harmonic oscillator) may be obtained from the free Schr\"odinger equation
 by the time reparametrization
$t(u)=\tan \omega u$ for which the Schwarzian derivative is a constant,  $\Theta(t)=2\omega^2$ (see
formulas in Proposition 1.5).
\item[(iv)] $V=k/r^2$ (inverse-square potential), corresponding to the operator $-2\II\partial_{\theta}-\partial_x^2+
\frac{x^2}{4}+kx^{-2}$ (harmonic oscillator with added inverse-square potential) in the Fourier coordinate. The operator is not in ${\cal S}_{\le 2}^{aff}$, but the (time-independent) inverse-square potential is interesting
in that this is the only potential left invariant by  all transformations $V(t,r)\to \phi'(t) V(\phi(t),r\sqrt{\phi'(t)})$ (see
formulas in Proposition 1.5). So this equation is invariant by the kernel of the Schwarzian derivative, i.e. by the  homographic transformations.
\end{itemize}


\section{Monodromy of time-dependent Schr\"odinger operators of non-resonant types  and Ermakov-Lewis invariants}


We 'solve' in this section all Schr\"odinger operators in ${\cal S}_{\le 2}^{aff}$ of class (i), (ii) or (iii) by
using the Ermakov-Lewis invariants, to be introduced below. Since any such operator is conjugate to an operator of the type
$-2\II\partial_{\theta}-\partial_x^2+V_2(\theta)x^2+\gamma$ ($\gamma$ constant), and $\gamma$ corresponds to a simple
energy shift, we shall implicitly assume that the potential is simply quadratic ($V_1=V_0=0$).

 Lemma 3.1.2 and Proposition 3.1.4  yield
 explicitly an evolution operator $U(\theta,\theta_0)$, i.e.
a unitary operator on $L^2(\R)$ which gives the evolution of the solutions of the Schr\"odinger equation from time
$\theta_0$ to time $\theta$. This operator gives the unique solution to the Cauchy problem and allows to compute
the (exact) Berry phase. The arguments in Lemmas 3.1.1, 3.1.2 and Proposition 3.1.4 are reproduced
from the article of Lewis and Riesenfeld (\cite{LewRie}). Unfortunately this method  gives the monodromy
only in the {\it elliptic} case (i.e. for operators of class (i) with $\alpha>0$). So we generalize their invariants
to the hyperbolic and unipotent case; the invariant we must choose in order to be able to compute the
monodromy  is not a harmonic oscillator any more, but an operator with absolutely continuous spectrum. Nevertheless,
it turns out that there does exist a phase operator, given in terms of the (possibly regularized) integral
$\int_0^{2\pi} \frac{d\theta}{\xi(\theta)}$ for a certain stabilizer $\xi$ of the quadratic part of the potential. The key
point in order to get the whole picture is to make the bridge between Kirillov's results and the Ermakov-Lewis invariants.

\subsection{Ermakov-Lewis invariants and Schr\"odinger-Virasoro invariance}

Let $H=\half(-\partial_x^2+V_2(\theta)x^2)$ be the (quantum) Hamiltonian corresponding to a time-dependent harmonic oscillator. The evolution of the wave function $\psi(\theta,x)$ is
given by: $\II\partial_{\theta}\psi(\theta,x)=H\psi(\theta,x)$, or $D\psi=0$
where $D=-2\II \partial_{\theta}+2H=-2\II\partial_{\theta}-\partial_x^2+V_2(\theta)x^2$.

The  {\it Ermakov-Lewis dynamical invariants} were invented in order to find the
solutions of the above equation. The idea is simple. Suppose $I(\theta,x)$ is a
{\it time-dependent} hermitian operator
of the form $\sum_{j=0}^N I_j(\theta,x)\partial_x^j$ which is an invariant of the motion, i.e. $\frac{d}{dt}I=\partial_t I+\frac{1}{\II} [I,H]=0$. Suppose also that, for every
fixed value of $\theta$,
$I(\theta,x)$ (defined on an appropriately defined dense subspace of $L^2(\R,dx)$, for
instance on the space of test functions) is essentially self-adjoint and has a purely point
spectrum. For simplicity, we shall assume that all multiplicities are one, and that
one may choose normalized eigenvectors  which depend regularly on $\theta$, namely,
\BEQ I(\theta,x)h_n(\theta,x)=\lambda_n(\theta) h_n(\theta,x) \EEQ
and $\int_{\R} |h_n(\theta,x)|^2 \ dx=1$.
 The fact that $I$ is an invariant of the motion implies
by definition that $I\psi$ is a solution of the Schr\"odinger equation if $\psi$ is.
The following lemma shows how to solve the Schr\"odinger equation by means of the
invariant $I$:

{\bf Lemma 3.1.1 (see \cite{LewRie})}
{\em

\begin{enumerate}
\item
The eigenvalues $\lambda_n(\theta)$ are constants, i.e. they do not depend on  time.
\item
{\em If} $n\not=m$, then $ \langle h_m(\theta),(\II\partial_{\theta}-H)h_n(\theta) \rangle=0$. 
\end{enumerate}
}

{\bf Proof.}

\begin{itemize}

\item[(i)]

Applying the invariance property $\frac{\partial I}{\partial \theta}+\frac{1}{\II}
[I(\theta),H(\theta)]=0$ to the eigenvector $h_n(\theta)$ yields
$$\frac{\partial I}{\partial\theta} h_n(\theta)+\frac{1}{\II}(I(\theta)-\lambda_n(\theta))H(\theta)h_n(\theta)=0.$$
Taking the scalar product with  $h_m(\theta)$ gives a first equation,
\BEQ
 \langle h_m(\theta),\frac{\partial I}{\partial\theta} h_n(\theta) \rangle +\frac{1}{\II}
(\lambda_m(\theta)-\lambda_n(\theta)) \langle h_m(\theta),H(\theta)h_n(\theta) \rangle=0.
\label{Lewis1}
\EEQ

The eigenvalue equation $I(\theta)h_n(\theta)=\lambda_n(\theta)h_n(\theta)$ gives
after time differentiation a second equation, namely
\BEQ \frac{\partial I}{\partial \theta} h_n(\theta)+(I(\theta)-\lambda_n(\theta))\dot{h}_n(\theta)=\dot{\lambda}_n(\theta)h_n(\theta). \label{Lewis2} \EEQ

Combining equations (\ref{Lewis1}) and (\ref{Lewis2}) for $n=m$ yields $\dot{\lambda}_n(\theta)=0$.

\item[(ii)]

Combining this time equations (\ref{Lewis1}) and (\ref{Lewis2}) for $n\not=m$ yields the
desired equality.
\end{itemize}

The above Lemma shows that one may {\it choose} eigenvectors $h_n(\theta)$ that satisfy
the Schr\"odinger equation by multiplying them by an appropriate time-dependent phase, 
which is the content of the following Lemma.

{\bf Lemma 3.1.2}

{\em
Let, for each $n$,  $\alpha_n(\theta)$ be a solution of the equation
\BEQ  \frac{d\alpha_n}{d\theta}=\langle h_n(\theta),\quad (\II  \partial_{\theta}-H)
h_n(\theta) \rangle. \label{phaseequation}\EEQ
Then the gauge-transformed eigenvectors for the invariant I
\BEQ \tilde{h}_n(\theta)=e^{\II\alpha_n(\theta)} h_n(\theta)\EEQ
are solutions of the Schr\"odinger equation.

In other words, the general solution of the Schr\"odinger equation is:
\BEQ \psi(\theta):=\sum_n c_n e^{\II \alpha_n(\theta)} h_n(\theta)\EEQ
where $c_n$ are constant (time-independent) coefficients.
}

Let us specialize to the case when $H$ is a time-dependent harmonic oscillator as above,
i.e. $H=\half(-\partial_x^2+V_2(\theta)x^2)$. A natural idea is to assume the following
Ansatz
$$I(\theta)=\half\left[ -b(\theta)\partial_x^2+a(\theta)x^2-\II c(\theta)(x\partial_x+\partial_x x)\right].$$  This problem has a unique family of non-trivial solutions:

{\bf Definition 3.1.3 (Pinney-Milne equation)}

{\em

The non-linear equation
\BEQ \ddot{\zeta}+f(\theta)\zeta-\frac{K}{\zeta^3}=0 \label{Pinney}\EEQ
$(K>0)$ is called a Pinney-Milne equation. If $K=1$, then we shall say that  (\ref{Pinney}) is a {\em normalized} Pinney-Milne equation.
}

Of course, every Pinney-Milne equation can easily be normalized by multiplying the function by the constant
factor $K^{1/4}$.

The following Proposition summarizes results due to Lewis and Riesenfeld (see \cite{LewRie}).

{\bf Proposition 3.1.4 (Ermakov-Lewis invariants for time-dependent harmonic oscillators)}

{\em
\begin{enumerate}
\item
The second-order operator ${\cal EL}(\zeta^2)$

\BEQ {\cal EL} (\zeta^2)(\theta)=\half\left[ \frac{x^2}{\zeta^2}+(\II \zeta(\theta)\partial_x+
\dot{\zeta}(\theta) x)^2\right] \EEQ

is an invariant of the time-dependent harmonic oscillator $-2\II \partial_{\theta}-\partial_x^2+V_2(\theta)x^2$ provided
 $\zeta$ is a solution of the following normalized {\it Pinney-Milne equation} :
\BEQ \ddot{\zeta}+V_2(\theta)\zeta-\frac{1}{\zeta^3}=0.\EEQ

Setting $\xi=\zeta^2$, one may also write equivalently
\BEA
 {\cal EL}(\xi)(\theta)&=& \frac{1}{2\xi} \left[ x^2+(\II \xi\partial_x+\half \dot{\xi} x)^2\right] \nonumber\\
 &=& \frac{1}{2\xi} \left[-\xi^2 \partial_x^2 +(1+\frac{\dot{\xi}^2}{4})x^2+ \frac{\II}{2}\xi \dot{\xi}
(x\partial_x+\partial_x x) \right].\EEA

\item

Set $$a(\theta):=\frac{1}{\sqrt{2}} \left[ \frac{x}{\zeta(\theta)}- (\zeta(\theta)\partial_x+\II \dot{\zeta}(\theta)x) \right]$$
and
 $$a^*(\theta)=\frac{1}{\sqrt{2}} \left[ \frac{x}{\zeta(\theta)}+ (\zeta(\theta)\partial_x+\II \dot{\zeta}(\theta)x) \right] $$
(formal adjoint of the operator $a(\theta)$). Then
\BEQ {\cal EL}(\xi)(\theta)=a^*(\theta)a(\theta)+\half.\EEQ
In other words, for every fixed value of $\theta$, the operators $a(\theta),a^*(\theta)$ play 
the roles of an annihilation, resp. creation operator for the (time-dependent) harmonic oscillator ${\cal EL}(\xi)$.

\item

The normalized ground state of the operator $a(\theta)$ is
\BEQ h_0(\theta)=\frac{1}{2\sqrt{\pi}} \frac{1}{\sqrt{\xi(\theta)}} \exp \left(
 \left(-\half \frac{1}{\xi(\theta)} + \frac{\II}{2} (\dot{\xi}/\xi)(\theta) \right) x^2 \right).\EEQ

\item
The solutions of equation (\ref{phaseequation}) giving the phase evolution of the solutions of the Schr\"odinger equation are given by 
\BEQ \alpha_n(\theta)=-(n+\half) \int^{\theta} \frac{d\theta'}{\xi(\theta')} \EEQ
provided one chooses the time-evolution of the eigenstates $h_n$ by setting
\BEQ \langle h_n, \partial_{\theta} h_n\rangle= \frac{\II}{2} (n+\half) (\zeta \ddot{\zeta}-\dot{\zeta}^2).\EEQ 

\end{enumerate}
}

The above  choice for the time-evolution of the eigenstates appears natural if one requires the standard
lowering and raising relations $a(\theta)h_n(\theta)=n^{\half} h_{n-1}(\theta)$, $a^*(\theta)h_n(\theta)=
(n+1)^{\half} h_{n+1}(\theta)$. Then computations show that
\BEQ \langle h_n, \partial_{\theta}h_n\rangle=\langle h_0,\partial_{\theta} h_0\rangle+\II \frac{n}{2}
(\zeta\ddot{\zeta}-\dot{\zeta}^2).\EEQ
Hence there only remains  to choose the time-evolution of the ground-state $h_0$. This particular choice leads to 
the $(n+\half)$-factor typical of the spectrum of the harmonic oscillator. Note that the $h_n(\theta)$ do {\it not} satisfy
the gauge-fixing condition typical of the adiabatic approximation (see A. Joye \cite{Joy92} for instance). But this phase
choice leads to a nice interpretation of the phases $\alpha_n$ (up to a constant) as a {\it canonical coordinate} conjugate
to the classical invariant ${\cal EL}_{cl}$ (see Lemma 3.2.2 below) for the corresponding classical problem, in
the generalized symplectic formalism for which  time is a coordinate, so that the problem becomes
autonomous  (see 
Lewis-Riesenfeld \cite{LewRie}; see also section 4 for the symplectic formalism). Also, as mentioned in the introduction, 
the natural time-scale (both for the classical and the
quantum problem) is $\tau(\theta):=\int^{\theta} \frac{du}{\xi(u)}.$

\bigskip

The connection with the preceding sections is given by the following classical lemma
(see \cite{MagWin}, chap. 3), which is an easy corollary of Proposition 2.3.2:

{\bf Lemma 3.1.5}

{\em
\begin{enumerate}
\item

Let $\xi$ be a (non-necessarily periodic) solution of  the  equation
\BEQ \half \xi'''+2u\xi'+u'\xi=0, \EEQ so that $\xi$ stabilizes $\partial^2+u$. 
 Then
$\zeta:=\sqrt{\xi}$ is a solution of the Pinney-Milne equation
\BEQ \zeta''+u(\theta)\zeta-\frac{I_u(\xi)/2}{\zeta^3}=0 \EEQ
where $I_u(\xi):=\xi \xi''-\half \xi'^2+2u\xi^2$ is the constant defined in Prop. 2.3.2 (2).

In particular, if $\xi=\psi_1^2+\psi_2^2$, where $(\psi_1,\psi_2)$ is a basis of solutions of the Hill equation
$(\partial^2+u)\psi=0$,  and $\zeta=\sqrt{\xi}$, then
\BEQ \zeta''+u(\theta)\zeta-\frac{W^2}{\zeta^3}=0 \label{PMW} \EEQ
where $W:=\psi_1\psi'_2-\psi'_1\psi_2$ is the Wronskian of the two solutions.

\item Consider $\xi\in Stab_u$ such that $\zeta=\sqrt{\xi}$ satisfies the Pinney-Milne equation (\ref{PMW}),
and a time-reparametrization $\phi$. Then $\tilde{\xi}:=\phi'^{-1}\ .\  \xi\circ\phi$ is a stabilizer of
$\partial^2+\tilde{u}:=\phi_*(\partial^2+u)$ and $\tilde{\zeta}:=\sqrt{\tilde{\xi}}$ satisfies the
transformed Pinney-Milne equation
$$\tilde{\zeta}''+\tilde{u} \tilde{\zeta}-\frac{W^2}{\tilde{\zeta}^3}=0$$ 
for the same constant $W$.
\end{enumerate}
}

The interesting point now is that one can choose the Ermakov-Lewis invariant in such a way that the 
invariant associated to the image of the time-dependent harmonic oscillator $D$ by a time reparametrization
(through the representation $\sigma_{1/4}$) is its image by a very natural transformation (essentially,
by the corresponding change of coordinates). This provides an elegant, natural explanation for the  complicated-looking
phase appearing in the formulas for $\sigma_{1/4}$. 

{\bf Theorem 3.1.6}

{\em
Let $D:=-2\II \partial_{\theta}-\partial_x^2+V_2(\theta)x^2$ be a time-dependent harmonic oscillator, $\zeta$
satisfy the Pinney equation $\zeta''+V_2 \zeta-\frac{1}{\zeta^3}=0$, and ${\cal EL}(\zeta^2)=\half\left[
\left(\frac{x}{\zeta}\right)^2+(\II\zeta \partial_x+\zeta' x)^2\right]$ be the associated Ermakov-Lewis
invariant. 

Let $\phi\in \Diff_+(\R/2\pi\Z)$ be a time-reparametrization and $\tilde{V}_2$ be the image of $V_2$ through
$\phi$, defined by $\sigma_{1/4}(\phi).D=-2\II \partial_{\theta}-\partial_r^2+\tilde{V}_2(\theta)x^2.$

Then:

\begin{enumerate}
\item
$\tilde{\zeta}:=(\phi'\circ\phi^{-1})^{\half} \ .\ \zeta\circ\phi^{-1}$ satisfies the transformed Pinney
equation $\tilde{\zeta}''+\tilde{V}_2 \tilde{\zeta}-\frac{1}{\tilde{\zeta}^3}=0.$
\item
Consider the transformed Ermakov-Lewis invariant
\BEQ \widetilde{{\cal EL}}(\tilde{\zeta}^2)(x):=\half \left[
\left(\frac{\tilde{x}}{\tilde{\zeta}}\right)^2+(\II\tilde{\zeta} \partial_{\tilde{x}}+ \frac{d \tilde{\zeta}}{d\tilde{\theta}}
 \tilde{x})^2\right]
\EEQ
where $(\tilde{\theta},\tilde{x})=(\phi(\theta),x\sqrt{\phi'(\theta)})$ are the transformed coordinates.

Then
\BEQ \widetilde{{\cal EL}}(\tilde{\zeta}^2)=\pi_{1/4}(\phi){\cal EL}(\zeta^2) \pi_{1/4}(\phi)^{-1}.\EEQ
In particular, $\widetilde{{\cal EL}}(\tilde{\zeta}^2)$ is an Ermakov-Lewis invariant for $\sigma_{1/4}(\phi)D$.
\end{enumerate}
}

{\bf Proof.}

1. follows from Lemma 3.1.5 (2).  This implies that $\widetilde{{\cal EL}}(\tilde{\zeta}^2)$ is an Ermakov-Lewis
invariant for $\sigma_{1/4}(\phi)\ .\ D$. Supposing one has proved that $ \widetilde{{\cal EL}}(\tilde{\zeta}^2)$
is  the conjugate of ${\cal EL}(\zeta^2)$ by $\pi_{1/4}(\phi)$, then it follows once again
 that $\widetilde{{\cal EL}}(\tilde{\zeta}^2)$ is an invariant for $\sigma_{1/4}(\phi)\ .\ D$ since
$$\left(\sigma_{1/4}(\phi)\ .\  D\right) \widetilde{{\cal EL}}(\tilde{\zeta}^2) - \widetilde{{\cal EL}}(\tilde{\zeta}^2) 
\left(\sigma_{1/4}(\phi)\ .\  D\right)
=\phi'\  \pi_{1/4}(\phi) D\ .\  {\cal EL}(\zeta^2)\ .\   \pi_{1/4}(\phi)^{-1}-\pi_{1/4}(\phi)\ .\ {\cal EL}(\zeta^2)\ .\ 
 \phi' D\ \pi_{1/4}(\phi)^{-1} =0$$
(the function of time $\phi'$ commutes with the operator ${\cal EL}(\zeta^2)$).

So all there remains to show is that  $ \widetilde{{\cal EL}}(\tilde{\zeta}^2)$ is indeed conjugate
to ${\cal EL}(\zeta^2)$. This is actually true for both terms appearing inside parentheses in the expression
for the Ermakov-Lewis invariant (and trivial for the first one). Set ${\cal E}=\II \zeta \partial_x+
\zeta' x$ and $\tilde{\cal E}=\II\tilde{\zeta} \partial_{\tilde{x}}+ \frac{d \tilde{\zeta}}{d\tilde{\theta}}
 \tilde{x}$. Then a simple computation shows that
$$\tilde{\cal E}=\II \zeta \partial_x+x\zeta'+\half x\frac{\phi''}{\phi'}\zeta.$$

On the other hand,
\BEA
&& (\pi_{1/4}(\phi) {\cal E} \pi_{1/4}(\phi)^{-1}) \psi(\tilde{\theta},\tilde{x}) = (\phi'(\theta))^{-1/4}
e^{\frac{1}{4} \II \frac{\phi''(\theta)}{\phi'(\theta)} x^2} {\cal E} \pi_{1/4}(\phi)^{-1} \psi(\theta,x) \nonumber\\
&& = \half   (\phi'(\theta))^{-1/4}
e^{\frac{1}{4} \II \frac{\phi''(\theta)}{\phi'(\theta)} x^2} (\II \zeta(\theta)\partial_x+\zeta'(\theta)x)
\left( \phi'(\theta)^{1/4} e^{-\frac{1}{4} \II \frac{\phi''(\theta)}{\phi'(\theta)} x^2} \psi(\phi(\theta),
x\sqrt{\phi'(\theta)}) \right) .
\EEA

Hence $\pi_{1/4}(\phi){\cal E}\pi_{1/4}(\phi)^{-1}=\tilde{\cal E}.$ 
\hfill \eop

\vskip 3 cm

We now want to be able to write the general solution of the Schr\"odinger equation as
$$\psi(\theta)=\int_{\Sigma} e^{\II \alpha_k(\theta)} c_k h_k(\theta) d\sigma(k)$$
(for some spectral measure $\sigma$ on a set $\Sigma$, a discrete measure in the case studied by Lewis and Riesenfeld) with {\it periodic} eigenstates $h_k$
and a phase $\alpha_k$ with {\it periodic derivative, i.e. given by integrating a periodic function}, so that
$$\psi(\theta+2\pi)=\int e^{\II \lambda_k} e^{\II \alpha_k(\theta)} c_k h_k(\theta) d\sigma(k)$$
where the $\lambda_k:=\alpha_k(\theta+2\pi)-\alpha_k(\theta)$
 are {\it constants} and measure the rotation of the eigenstates $h_k$ after a time $2\pi$. Then the
monodromy operator is unitarily equivalent to the multiplication operator  $f(k)\to f(k)e^{\II\lambda_k}$ on
$L^2(\Sigma,d\sigma)$.

Consider any Schr\"odinger operator with quadratic potential $V_2(\theta)x^2$ and an associated non-zero vector field $\xi\in Stab(V_2)$
as before. (We postpone the discussion of 'resonant' operators (classes (i)bis and (iii)bis) to the next section.)
It turns out that the eigenstates $h_k$ and the measure $\sigma$ can be taken as  the (possibly generalized) eigenfunctions and spectral measure of one of the three following 'model' operators $H$, depending on the {\it sign of the invariant} $I_u(\xi)$:

\begin{itemize}
\item[(i)]  ($I_u(\xi)>0$) : take  for $H$ the standard harmonic oscillator 
 $$H=-\half(\partial_x^2-a^2 x^2) \quad (a\in\R);$$  this case corresponds
to harmonic oscillators of type (i), i.e. Schr\"odinger operators of type (i) conjugate to $-2\II \partial_{\theta}-\partial_x^2+a^2 x^2$
with $a^2>0$;
\item[(ii)]  $(I_u(\xi)<0)$ : take for $H$ the 'standard harmonic repulsor'
   $$H=-\half(\partial_x^2+a^2 x^2) \quad (a\in\R); $$  this case corresponds
to harmonic repulsors of type (i), i.e. operators of type (i) conjugate to  $-2\II \partial_{\theta}-\partial_x^2-a^2 x^2$ ($-a^2<0$),
and operators of type (ii);
\item[(iii)]   $(I_u(\xi)=0)$ : take for $H$ the usual one-dimensional Laplacian,
  $$H=-\half\partial_x^2;$$ this case corresponds to operators of type
(i) conjugate to the free Schr\"odinger operator $-2\II \partial_{\theta}-\partial_x^2$, and operators of type (iii).
\end{itemize}

Note that this classification is equivalent to the classification of the (conjugacy classes of) monodromy matrices
for the associated Hill operators $\partial_{\theta}^2+V_2(\theta)$ into elliptic, hyperbolic and unipotent elements.

The next section circumvents the spectral analysis technicalities by solving the associated classical problem. The essentials
for understanding  the  (operator-valued) monodromy
for the quantum problem are already contained in the study of the ($SL(2,\R)$-valued) monodromy of the ordinary differential
equation $\ddot{x}=-V_2(\theta) x$, so we found this short digression convenient for the reader. Then we study the spectral
decomposition of the above model operators. Finally, we solve the quantum problem for a quadratic potential $V_2(\theta)x^2$ and
compute the monodromy operator. The general case $D\in {\cal S}^{aff}_{\le 2}$ may be reduced to the quadratic case
$D\in {\cal S}^{aff}_{2}$ after applying some transformation in SV, except for the operators of type (i)bis and (iii)bis which will be treated in the last section.


\subsection{Solution of the associated classical problem}


The associated classical problem (obtained for instance as the lowest-order term in $\hbar$ in the usual
semi-classical expansion) is a Hill equation.

{\bf Definition 3.2.1 (classical problem)}

{\em
Let $H$ be the classical hamiltonian $H=\half(p^2+V_2(\theta)x^2)$. 
}

The asociated motion in phase space reads $\dot{x}=\partial_p H=p$, $\dot{p}=-\partial_x H=-V_2$, which is
equivalent to the Hill equation  $(\partial_{\theta}^2+V_2)x(\theta)=0$.

{\bf Lemma 3.2.2}

{\em
\begin{enumerate}
\item
Suppose $V_2$ is of type I with $\alpha\not =0$ or of type  II, and choose $\xi\in \Stab V_2$ so that $I_u(\xi)=2$ ($\xi$ is real in the elliptic case and
purely imaginary in the hyperbolic case). Then
\BEQ {\cal EL}_{cl}(\xi)(x):=\half\left[ \frac{x^2}{\xi}+\xi(\dot{x}-\half \frac{\dot{\xi}}{\xi} x)^2\right] \EEQ
is an invariant of the motion.
\item
Suppose $V_2$ is of type I with $\alpha=0$ or of type III (so that the associated monodromy is unipotent), and take any
$\xi\in \Stab V_2$, $\xi\not=0$. Then
\BEQ {\cal EL}_{cl}(\xi)(x):=\half \left[ \xi(\dot{x}-\half \frac{\dot{\xi}}{\xi} x)^2\right] \EEQ
is an invariant of the motion.
\end{enumerate}
}

{\bf Proof.} Simple computation (${\cal EL}_{cl}$ may be obtained from the quantum Ermakov-Lewis invariant by letting $\hbar$
go to zero). \hfill \eop

Assuming $V_2$ is elliptic, i.e. of type I with $\alpha>0$, one may choose $\xi>0$. Then the equation ${\cal EL}_{cl}(\xi)(x)=C$,
$C$ constant is equivalent to $\left(\frac{dz}{d\tau}\right)^2+z^2=C$ after the function- and time-change
$\tau(\theta)=\int^{\theta} \frac{d\theta'}{\xi(\theta')}$, $x(\theta)=\xi^{\half}(\theta) z(\tau(\theta))$, with
obvious solutions $\cos\tau$, $\sin\tau$. Hence a basis of solutions of the equation of motion is given by
\BEQ x_1(\theta)=\xi^{\half}(\theta) \cos \int^{\theta} \frac{d\theta'}{\xi(\theta')}, \quad
x_2(\theta)=\xi^{\half}(\theta) \sin \int^{\theta} \frac{d\theta'}{\xi(\theta')}    \label{x1x2} \EEQ
Assume for instance  that $\dot{\xi}(0)=0$, and choose $\int^{\theta} \frac{d\theta'}{\xi(\theta')}=\int_0^{\theta} \frac{d\theta'}{\xi(\theta')}$.
 Then $\left(\begin{array}{c} x_1 \\ x_2 \end{array}\right) (2\pi)=
\left(\begin{array}{cc} \cos T & -\sin T\\ \sin T & \cos T\end{array}\right) \ .\ \left(\begin{array}{c} x_1 \\ x_2 \end{array}
\right) (0)$ with $T=\int_0^{2\pi} \frac{d\theta'}{\xi(\theta')}$. Hence the eigenvalues of the monodromy matrix are given
by $\pm \II T$.

In the hyperbolic case (type I with $\alpha<0$, or type II), $\xi:=\II \eta$ is purely imaginary. The above formulas (\ref{x1x2}) give
solutions of the Hill equation on 
either side of any zero of $\xi$ (note that the normalization $I_u(\xi)=2$ implies $\xi(\theta)\sim_{\theta\to\theta_0}
\pm 2\II (\theta-\theta_0)$ near any zero, so that (\ref{x1x2}) defines a continuous function, as should be of course), but
the easiest way to define the solutions $x_1,x_2$ globally is to use a deformation of contour. One may always assume that
$\xi$ is analytic on some complex neighbourhood of $\R$ (it is conjugate by a time-reparametrization to some
$u_{n,\alpha}$ which is entire, see Prop. 2.2.3). Define a contour $\Gamma$ from $0$ to $2\pi$ which avoids the zeros of $\xi$
by going around them  along half-circles of small radii centered on the real axis. This time (see discussion
in section 2.2), the half-circles must be chosen alternatively
in the upper- and lower-half planes so that $\Re\xi(z)\ge 0$ on $\Gamma$. Then $\left(\begin{array}{c} x_1 \\ x_2\end{array} \right) (2\pi)=
\left(\begin{array}{cc} \cos T & -\sin T\\ \sin T & \cos T\end{array}\right) \ .\ \left(\begin{array}{c} x_1 \\ x_2 \end{array}
\right) (0)$ as before,  with $T=\int_{\Gamma} \frac{d\theta'}{\xi(\theta')}$. Mind that $T$ is purely imaginary this time. 

Finally, in the unipotent case (type I with $\alpha=0$, or type III), normalize $\xi$ by setting for instance
$\xi(0)=\II$, $\dot{\xi}(0)=0$, so that $\xi$ is purely imaginary. The same function- and time-change yields
 $\left(\frac{dz}{d\tau}\right)^2=C$, hence a natural
basis of solutions is given by $x_1(\theta)=\xi^{\half}(\theta),x_2(\theta)=\xi^{\half}(\theta) \int_0^{\theta}
 \frac{d\theta'}{\xi(\theta')}$. To get globally define solutions, one avoids the double zeros of $\xi$ by drawing half-circles in the upper
half-plane. Then the monodromy matrix is $\left(\begin{array}{cc} 1 & T\\ 0& 1\end{array}\right)$, with
$T=\int_{\Gamma}\frac{d\theta'}{\xi(\theta')}=\int_{\Gamma}\frac{d\theta'}{x_1^2(\theta')}.$

\bigskip


\subsection{Spectral decomposition of the model operators}


We shall need below the spectral decomposition of the three model operators $-\half(\partial_x^2 - a^2 x^2)$, $-\half(\partial_x^2+a^2 x^2)$,
 $-\half\partial_x^2$ introduced above. They are essentially self-adjoint on $C_0^{\infty}(\R)$ by the classical Sears theorem (see
\cite{BerShu}, Theorem 1.1 chap. 2 for instance), so the spectral theorem applies. The first operator has a  pure point spectrum,
while the second and the third have an absolutely continuous spectrum. Note that $-\half\partial_x^2$ is non-negative, while the spectrum
of $-\half(\partial_x^2+a^2 x^2)$ is the whole real line, as the following Lemma proves.

{\bf Lemma 3.3.1}

{\em
\begin{enumerate}
\item (elliptic case)

The spectral decomposition of $L^2(\R)$ for the operator $-\half(\partial_x^2-a^2 x^2)$ is given by
\BEQ L^2(\R)=\oplus_{n\ge 0} L^2_{a(n+\half)} \EEQ
where $L^2_{a(n+\half)}$ is one-dimensional, generated by the normalized Hermite functions $Ca^{1/4} e^{-ax^2/2} 
He_n(x\sqrt{a})$ for some constant $C$ (see \cite{Abra84} for
the notations and normalization).

\item (hyperbolic case)

Set, for $\lambda\in\R$,
\BEA \psi_{\lambda}^{\pm}(x) & := &\left(\frac{2}{a}\right)^{1/4} e^{\lambda/8a} e^{-\II ax^2/2}  \nonumber\\
&& \left[ \frac{1}{\Gamma(\frac{3}{4}+\frac{\II \lambda}{4a})} \ _1 F_1(\frac{1}{4}(1+\frac{\II \lambda}{a}),\half;\II ax^2)
\pm \frac{2\sqrt{a}}{\Gamma(\frac{1}{4}+\frac{\II \lambda}{4a})} e^{\II\pi/4} x \ .\ _1 F_1(\frac{1}{4}(3+\frac{\II \lambda}{a}),
\frac{3}{2};\II ax^2) \right] \nonumber\\
\EEA
where $_1 F_1$ is the usual confluent hypergeometric function. Then $H\psi_{\lambda}^{\pm}=\lambda \psi_{\lambda}^{\pm}$, and
the $(\psi_{\lambda}^{\pm},\lambda\in\R)$ form a complete orthonormal system of generalized eigenfunctions of the operator
$H=-\half(\partial_x^2+a^2 x^2)$, so that any function $f\in L^2(\R)$ decomposes uniquely as
\BEQ f(x)=\int_{\R} \psi_{\lambda}^+(x) \overline{g^+(\lambda)} \ d\lambda+\int_{\R} \psi_{\lambda}^-(x) \overline{g^-(\lambda)} \ d\lambda \EEQ
with $g^{\pm}(\lambda)=\int_{\R} f(x) \overline{\psi_{\lambda}^{\pm}(x)} \ dx.$ In particular, the following Parseval identity holds,\BEQ \int_{\R} |f(x)|^2\ dx=\int_{\R} |g^+(\lambda)|^2 \ d\lambda+\int_{\R} |g^-(\lambda)|^2\ d\lambda.\EEQ

\item (unipotent case)

Set, for $\lambda> 0$, $\psi_{\lambda}^{\pm}(x)=e^{\pm \II x\sqrt{2\lambda}}$. Then $H\psi_{\lambda}^{\pm}=\lambda \psi_{\lambda}^{\pm}$ and the $\psi_{\lambda}^{\pm}$, $\lambda>0$, form a complete orthonormal system of generalized eigenfunctions of the operator
$H=-\half\partial_x^2$, with the usual Parseval-Bessel identity.
\end{enumerate}
}

{\bf Proof.}

(i) is classical and (iii) is straightforward by Fourier inversion and the usual Parseval-Bessel identity. Case (ii) is less
common, though it can certainly be found somewhere in the literature. Let us explain briefly how to obtain its spectral decomposition
for $a=1$. The easiest way is to remark that $H=A\Lambda A^{-1}$ where $\Lambda=\frac{\II}{2} (x\partial_x+\partial_x x)
=\II(x\partial_x+\half)$ and $A$ is the image of the rotation matrix $\left(\begin{array}{cc} \cos\pi/4 & -\sin \pi/4 \\ \sin\pi/4 & \cos \pi/4
\end{array}\right)$ by the metaplectic representation. The operator $A$ is unitary. Explicit formulas found
for instance in \cite{GuiSte} show that
\BEQ (Af)(x)=\II \sqrt{2} e^{\II \pi/4} e^{\II\pi x^2} \int_0^{\infty} e^{-\II \pi(x\sqrt{2}-y)^2} f(y)\ dy.\EEQ
As for the operator $\Lambda$, it is conjugate to $\II (\partial_y+\half)$ after the obvious change of variable
$x=\pm e^y$, hence its spectral decomposition is given by Fourier inversion on either half-lines, $\Lambda \phi_{\lambda}^{\pm}=\lambda \phi_{\lambda}^{\pm}$ ($\lambda\in\R$) with $\phi_{\lambda}^{\pm}(x)=x_{\pm}^{-\half-\II \lambda}$ constituting an
orthonormal basis of generalized eigenfunctions. Finally, $\psi_{\lambda}^{\pm}:=A\phi_{\lambda}^{\pm}$ may be obtained
 by applying the following formula (see \cite{GradRyzh})
$$ \int_0^{\infty} x^{\nu-1} e^{-\beta x^2-\gamma x}\ dx=(2\beta)^{-\nu/2} \Gamma(\nu) e^{\gamma^2/8\beta} D_{-\nu}(\frac{\gamma}{\sqrt{2\beta}})$$
($\Re \beta,\Re\nu>0$) where $D_{\nu}$ is a parabolic cylinder function, also given by
\BEQ D_{\nu}(z)=2^{\nu/2} e^{-z^2/4} \left\{ \frac{\sqrt{\pi}}{\Gamma(\half(1-\nu))} \ _1 F_1(-\frac{\nu}{2},\half;z^2/2)
-z\frac{\sqrt{2\pi}}{\Gamma(-\nu/2)} \ _1 F_1(\frac{1-\nu}{2},\frac{3}{2};z^2/2) \right\} \EEQ
(see \cite{Erd53}, 8.2. (4) p. 117).
\hfill \eop


\subsection{Monodromy of non-resonant harmonic oscillators  (elliptic case)}


We assume here that $D\in {\cal S}_{\le 2}^{aff}$ is of class (i) with $\alpha>0$. Then $D$ is conjugate by a transformation
in SV to an operator of the type $-2\II \partial_{\theta}-\partial_x^2+a^2 x^2+\gamma$ where $a>0$ and $\gamma$ is a constant.
Choose $\xi=\frac{1}{a}$ so that $\sqrt{\xi}$ satisfies a
normalized Pinney-Milne equation. Then Proposition 3.1.4 shows the following:

{\bf Theorem 3.4.1}
{\em

The solution of the Schr\"odinger equation
with  arbitrary  initial state
\BEQ \psi(0):=\sum_{n\ge 0} c_n h_n(0) \EEQ is given by
\BEQ \psi(\theta):=\sum_{n\ge 0} c_n e^{-\II (n+\half) a\theta-\II \gamma \theta/2} h_n(\theta). \EEQ
The monodromy operator is given by the 'infinite-dimensional' monodromy matrix $ M_D:=\diag(e^{\II \lambda_n},n\in\N)$,
with $\lambda_n:=-2\pi (n+\half)a- \pi\gamma$. 
}


\subsection{Monodromy of harmonic repulsors  (hyperbolic type)}


One assumes now that $D\in {\cal S}_{\le 2}^{aff}$ is either of class (i) with $\alpha<0$ or of class (ii).
Consider again the Ermakov-Lewis invariant
\BEQ {\cal EL}(\xi)=\frac{1}{2\xi} \left[ x^2+(\II \xi\partial_x +\half\dot{\xi}x)^2 \right]\EEQ
where one has assumed that  $\xi=\II \eta$ is {\it purely imaginary} this time, and $I_{V_2}(\xi)=2$. Note that
${\cal EL}(\II\eta)$ is {\it anti-hermitian}. Then
\BEQ \frac{{\cal EL}(\II\eta)-\II k}{\II\eta}=-\half\left[ \partial_x^2-\II \frac{\dot{\eta}}{\eta} x\partial_x+\frac{1-\frac{1}{4}\dot{\eta}^2}{\eta^2} x^2-\half \II \frac{\dot{\eta}}{\eta}+\frac{2k}{\eta} \right].\EEQ

Suppose $\psi_k\not=0$ is an eigenvector of the Ermakov-Lewis operator, ${\cal EL}(\II\eta)\psi_k=\II k\psi_k$. Then Proposition 2.1.4 implies that  $\tilde{\psi}_k:=\exp- \frac{\II}{4}\frac{\dot{\eta}}{\eta}x^2 \ .\ \psi_k$ is a generalized eigenfunction of the
model harmonic repulsor, namely
\BEQ -\half(\partial_x^2+\frac{ x^2}{\eta^2})\tilde{\psi}_k=\frac{k}{\eta} \tilde{\psi}_k.\EEQ

Hence:

{\bf Lemma 3.5.1}

{\em
\begin{enumerate}
\item
The equation $({\cal EL}(\II\eta)-\II k)\psi_k=0$ $(k\in\R)$ has two linearly  independent solutions,
 $$\psi^k_{even}(\theta,x)=\sqrt{2} (2\II\eta)^{1/4} e^{k/4}e^{\frac{\II}{4} \frac{\dot{\eta}}{\eta}x^2} e^{-\frac{\II}{2\eta}x^2} \ .\ \frac{1}{\Gamma(\frac{1}{4}+\II\frac{k}{2})} \ _1 F_1(\frac{1}{4}(1+2\II k),\half;
\frac{\II x^2}{\eta})$$ and
 $$\psi^k_{odd}(\theta,x)=2\sqrt{2} (2\II\eta)^{1/4} e^{k/4} e^{\frac{\II}{4} \frac{\dot{\eta}}{\eta}x^2} e^{-\frac{\II}{2\eta}x^2} \frac{1}{\Gamma(\frac{3}{4}+\II\frac{k}{2})} x\ .\  _1 F_1(\frac{1}{4}(3+2\II k),\frac{3}{2};
\frac{\II x^2}{\eta}).$$

The functions $((\psi^k_{even},\psi^k_{odd}), k\in\R)$ constitute a complete orthonormal system for the
operator ${\cal EL}(\II\eta).$
\item

One has
$$D \psi^k_{even}(x)=\left(\frac{2k}{\eta}-\II\frac{\dot{\eta}}{\eta}\right) \psi^k_{even}(x)$$
and
$$D \psi^k_{odd}(x)=\left(\frac{2k}{\eta}-2\II\frac{\dot{\eta}}{\eta}\right) \psi^k_{odd}(x).$$

Hence $x\to \frac{1}{\sqrt{\xi}} \exp\left(  k \int^{\theta} \frac{d\theta'}{\xi(\theta')}\right)
\psi^k_{even}(x)$ and $x\to \frac{1}{\xi} \exp\left(  k \int^{\theta} \frac{d\theta'}{\xi(\theta')}\right)
\psi^k_{odd}(x)$ are solutions of the Schr\"odinger equation.
\end{enumerate}
}

{\bf Proof.}

  1. is a direct application of Lemma 3.3.1, while 2. follows from an easy computation using the
confluent hypergeometric differential equation $z\frac{d^2}{dz^2}\  _1 F_1(a,c;z)+(c-z)\frac{d}{dz}\ _1 F_1(a,c,;z)
-a\ _1 F_1(a,c;z)=0.$ \hfill \eop

The eigenfunctions $\psi^k_{even}$, $\psi^k_{odd}$ depend analytically on $\xi$ for $\xi\in\C\setminus\R_-$.
If the operator $D$ is of type I (so that $\xi$ has no zero), say with $\gamma=0$,
 then the phase $\exp \left( k\int^{\theta}
\frac{d\theta'}{\xi(\theta')} \right)$
gives the monodromy. If $D$ is of type II, then one must resort to a deformation of contour in order
to avoid the singularities, as in the classical case, see section 3.2. Mind that the deformation of contour
may  change drastically  the  behaviour of the functions $\psi^k_{even}$, $\psi^k_{odd}$ for large $x$ or large $k$ (for
instance, $\psi^k_{even}$ and $\psi^k_{odd}$ become exponentially increasing for large $x$).
 Hence, in order to be able to follow the phase shift of the eigenfunctions
$\psi^k_{even}$, $\psi^k_{odd}$ along the contour $\Gamma$ without getting divergent integrals, it is better to assume to begin with that the 'Fourier transform' (with respect to the spectral decomposition of ${\cal EL}(\xi)$)
 of the solution has compact support. In other words, the solution of the Schr\"odinger equation with
initial state
$$\psi(0,x):=\int_{\R} \bar{c}_+(k) \psi^k_{even}(0,x)\ dk+ \int_{\R} \bar{c}_-(k) \psi^k_{odd}(0,x)\ dk$$
for $z\in\Gamma$  (complex time), where $c_+,c_-$ are assumed to be compactly supported, is given by
$$\psi(z,x)=\sqrt{\frac{\xi(0)}{\xi(z)}} \int_{\R} \bar{c}_+(k) e^{  k\int_0^z \frac{dz'}{\xi(z')} -\II \gamma\theta/2}  \psi^k_{even}(z,x)\ dk+ 
\frac{\xi(0)}{\xi(z)} \int_{\R} \bar{c}_-(k) e^{ k\int_0^z \frac{dz'}{\xi(z')}-\II \gamma\theta/2} \psi^k_{odd}(z,x)\ dk$$

An immediate corollary is:

{\bf Theorem 3.5.2}

{\em
Let $\psi(0)\in L^2(\R)$, with decomposition
\BEQ \psi(0,x):=\int_{\R} \bar{c}_+(k) \psi^k_{even}(0,x)\ dk+ \int_{\R} \bar{c}_-(k) \psi^k_{odd}(0,x)\ dk.\EEQ
Then the solution of any type (ii) Schr\"odinger equation with initial state $\psi(0)$ is given at time $\theta=2\pi$ by
\BEQ \psi(2\pi,x)=\int_{\R} \bar{c}_+(k) e^{ k T-\II\pi \gamma}  \psi^k_{even}(0,x)\ dk+
 \int_{\R} \bar{c}_-(k) e^{ kT-\II\pi \gamma} \psi^k_{odd}(0,x)\ dk \EEQ
where $T=\int_0^{2\pi} \frac{du}{\xi(u)}$ or $\int_{\Gamma} \frac{du}{\xi(u)}$ (depending on the
class of $V_2$), with $\Gamma$ chosen as in section 3.2, is purely imaginary.
The associated monodromy operator in ${\cal B}(L^2(\R),L^2(\R))$ is unitarily equivalent to the multiplication by the function
 $k\to e^{kT-\II\pi\gamma}$ with modulus one.
}


\subsection{Monodromy of non-resonant operators of unipotent type}


Suppose now $D\in {\cal S}_{\le 2}^{aff}$ is of class (i), $\alpha=0$ or (iii).
Then 
\BEQ {\cal EL}(\xi)(\theta):=\frac{1}{2\xi} \left[ (\II \xi\partial_x+\half \dot{\xi} x)^2 \right] \EEQ
($\xi\in\Stab_{V_2}$) is an invariant of $D$ (note the difference with respect to Proposition 3.1.4). Case (i), $\alpha=0$
is trivial, for it is conjugate to the free Schr\"odinger equation. So assume $D=-2\II\partial_{\theta}-\partial_x^2+V_2 x^2$ is of
class (iii). Take $\xi=\II\eta$ with $\eta\ge 0$ as in section 3.2. Then (if $k>0$)
\BEQ \frac{{\cal EL}(\xi)-\II k}{\xi}=-\half (\partial_x-\frac{\II}{2}\frac{\dot{\eta}}{\eta}x)^2 -\frac{k}{\eta}. \EEQ
So 
\BEQ \psi_{k,\pm}(x):=\exp \frac{\II}{4} \frac{\dot{\eta}}{\eta}x^2 \ . \ \exp \pm \II  \sqrt{\frac{2k}{\eta}} x \EEQ
constitute a complete orthonormal system for ${\cal EL}(\II \eta)$ (the same statement holds true for potentials of class (i), in which
 case $\eta=1$ and the exponential prefactor is trivial). A short computation shows that  
$$D\psi_{k,\pm}=\left(\frac{2k}{\eta}-\frac{\II}{2}\frac{\dot{\eta}}{\eta}\right)\psi_{k,\pm}.$$
Hence one has the following:

{\bf Theorem 3.6.1}

{\em

Let $\psi(0)\in L^2(\R)$, with decomposition
\BEQ \psi(0,x):=\int_{\R_+} \bar{c}_+(k) \psi_{k,+}(x)\ dk \ +\ \int_{\R_+} \bar{c}_-(k) \psi_{k,-}(x)\ dk. \EEQ
Then the solution of any type (i), $\alpha=0$ or type (iii) Schr\"odinger equation with initial state $\psi(0)$ is given at time $\theta=2\pi$ by
\BEQ \psi(2\pi,x)= \int_{\R_+} \bar{c}_+(k) e^{ k T-\II\pi \gamma} \psi_{k,+}(x) \ dk +  \int_{\R_+} \bar{c}_-(k)
e^{kT-\II\pi\gamma} \psi_{k,-}(x)\ dk, \EEQ
where $T=\int_0^{2\pi} \frac{du}{\xi(u)}$ or  $\int_{\Gamma} \frac{du}{\xi(u)}$, $\Gamma$ chosen as
in section 3.2 (depending on the
class of $V_2$) is purely imaginary.
The associated monodromy operator in ${\cal B}(L^2(\R),L^2(\R))$ is unitarily equivalent to the unitary
operator on $L^2(\R_+)$ given by the multiplication by the function $k\to e^{kT-\II\pi\gamma}$.
}


\section{Symplectic structures and general solution of the Schr\"odinger equation}


The general emphasis in this section is so to speak on the non-quadratic part of the potential, namely, on $V_0$ and $V_1$ if
$D=-2\II\partial_{\theta}-\partial_x^2+V_2(\theta)x^2+V_1(\theta)x+V_0(\theta)$. It contains somewhat loosely related results:
a definition of a three-dimensional invariant $(\xi,\del_1,\del_2)$; a generalization of the Ermakov-Lewis invariants to general
potentials; a symplectic structure on a space 'containing' ${\cal S}_{\le 2}^{aff}$ such that the SV-action becomes naturally
Hamiltonian; finally, the computation of the monodromy for the 'resonant' operators of type (i)bis, (iii)bis.

{\bf Definition 4.1}

{\em

We shall say that $D\in {\cal S}_{\le 2}^{aff}$ is of {\em generic type} if: $D$ is of class (i), $D$ conjugate
to $D_{\alpha,\gamma}=-2\II \partial_{\theta}-\partial_x^2+\alpha x^2+\gamma$ with $\alpha\not=n^2/4$, $n=0,1,\ldots$; or $D$ is of class (ii), $D$ conjugate to $D_{n,\alpha,\gamma}=-2\II \partial_{\theta}-\partial_x^2+
u_{n,\alpha}(\theta)x^2+\gamma$.

Denote by ${\cal S}_{\le 2,gen}^{aff}$ the set of operators of generic type; it is a disjoint union of $SV$-orbits.
}

Note (see Corollary 2.4.4) that the isotropy group of an operator $D$ of generic type is generated by ${\cal M}_1$ and some
${\cal L}_{\xi}+{\cal Y}_{f_1}+{\cal M}_{f_2}$ with $\xi\not=0$.

{\bf Definition 4.2}

{\em
Let $D=-2\II \partial_{\theta}-\partial_x^2+V_2(\theta)x^2+V_1(\theta)x+V_0(\theta)\in {\cal S}_{\le 2,gen}^{aff}$
be of generic type. 

Define:
\begin{itemize}
\item[(i)] $\xi(D)$ to be the unique (up to a sign) periodic vector field such that $\xi(D)\in \Stab_{V_2}$ 
and $I_{V_2}(\xi(D))= 2$ ($\xi$ real in the elliptic case, purely imaginary in the hyperbolic case);
\item[(ii)] $\del_1(D)$ to be the unique periodic function such that
\BEQ \ddot{\del_1}(D)+V_2 \del_1(D)=-\half (\dot{V}_1 \xi(D)+\frac{3}{2} V_1 \dot{\xi}(D)) ; \label{defdel1} \EEQ
\item[(iii)] $\del_2(D)$ to be the unique periodic function (up to a constant) such that
\BEQ \del_2(D)=-\half\int^{\theta} V_1(\theta') \del_1(D)(\theta') d\theta' -\half V_0 \xi(D) \label{defdel2} \EEQ
\end{itemize}
}

Observe that ${\cal L}_{\xi}+{\cal Y}_{\del_1}+{\cal M}_{\del_2}\in Lie(G_D)$ is indeed unique (up to
the addition of a constant times ${\cal M}_1$) as follows from Corollary 2.4.4. The ambiguity in the definition
of $\del_2$ may be solved by choosing for each SV-orbit an arbitrary base-point, an invariant $(\xi,\del_1,\del_2)$
for this base-point, and transforming $(\xi,\del_1,\del_2)$ covariantly by the adjoint action along the orbit. Some
non-local formulas fixing $\del_2$ more explicitly can probably be found, at least for potentials of type (i)
(see Lemma 2.4.5), but we shall not need them.

Another problem comes from the fact that the map $(V_2,V_1,V_0)\to(\xi,\del_1,\del_2)$ is not one-to-one (nor
onto).  Suppose one has some triple of functions $(\xi,\del_1,\del_2)$. Under some conditions that we shall not
write explicitly (depending on the class of the potential), $(\xi,\del_1,\del_2)$ is an invariant for some potential
$(V_2,V_1,V_0)$; the quadratic part $V_2$ is given (by definition) by $V_2=\frac{1}{2\xi^2}(2-\xi \ddot{\xi}+
\half \dot{\xi}^2)$. (Supposing $\xi$ has only a finite number of zeros, all of which are simple or double, one has
some rather straightforward conditions on the values of $\frac{d\xi}{d\theta}$ and $\frac{d^3 \xi}{d\theta^3}$ at the zeros of $\xi$ that ensure
that $\xi\in\Stab_{V_2}$ for some potential $V_2$).  But $V_1$ is not determined uniquely
 if $\xi$ does not vanish on the torus, since $\xi^{-3/2}$ is in the kernel of the operator
$\xi\partial+\frac{3}{2}\dot{\xi}$ (see formula (\ref{defdel1})). This can easily be explained by supposing
(by conjugating by some element $g\in SV$) that $D$ is the model operator $D=-2\II\partial_{\theta}-\partial_x^2+\alpha
x^2+\gamma$ ($\alpha$ generic). Then $\xi$ is proportional to the constant vector field ${\cal L}_1$ which commutes
with ${\cal Y}_1$, hence the invariant $(\xi,\del_1,\del_2)$ is left unchanged by space-translations, whereas
the operator $D$ (and also the generalized Ermakov-Lewis invariant defined in Theorem 4.4 below) is not. Hence
the vector invariant $(\xi,\del_1,\del_2)$ parametrizes Schr\"odinger operators of type (i) 'up to space-translations'.
On the other hand, the map $(V_2,V_1,V_0)\to(\xi,\del_1,\del_2)$ is one-to-one for operators of type (ii) (up
to a sign for $\xi$).  

It is not {\em a priori} self-evident that $\del_2$ defined by equation (\ref{defdel2}) is a periodic function.
Considering the 'inverse problem', i.e. supposing that the invariant $(\xi,\del_1,\del_2)$ is given, and supposing
$\xi$ does not vanish on the torus, one must also check that every choice for $V_1$ gives a function $\del_2$
which is periodic. This is the content of the following lemma:

{\bf Lemma 4.3}

{\em
One has:
\BEQ \frac{d}{d\theta} \left( \xi \frac{d}{d\theta}(\xi^{-\half}\del_1)\right)=-\half \frac{d}{d\theta}(\xi^{3/2}V_1)
-\xi^{-3/2} \del_1.\EEQ
This formula implies: $\int_0^{2\pi} \xi^{-3/2}\del_1 =0$; $\int_0^{2\pi} V_1\del_1=0$.
}

{\bf Proof.}

Using the invariant equations $\xi \ddot{\xi}-\half \dot{\xi}^2+2V_2\xi^2=2$ and $\ddot{\del}_1+V_2\del_1=-\half
(\dot{V}_1 \xi+\frac{3}{2} V_1 \dot{\xi})$, one obtains
$$\frac{d}{d\theta} (\xi^{\half} \dot{\del}_1)=\half \frac{d}{d\theta} (\xi^{-\half}\dot{\xi}\del_1)-\half
\frac{d}{d\theta}(\xi^{3/2}V_1)-\xi^{-3/2}\del_1,$$
hence the first equation, which implies immediately: $\int_0^{2\pi} \xi^{-3/2}(\theta)\del_1(\theta)\ d\theta=0$.
 Hence (considering
the inverse problem),\  {\em if} some
potential $V_1$ verifies $\int_0^{2\pi} V_1(\theta)\del_1(\theta)\ d\theta=0$ (so that $\del_2$ is well-defined),
then this is also true for all possible potentials $V_1$. Now, integrating the first equation, one gets
$$\xi \frac{d}{d\theta} (\xi^{-\half}\del_1)+\half \xi^{3/2}V_1=-\int^{\theta} \xi^{-3/2}(\theta')\del_1(\theta')
d\theta',$$
hence $$\xi^{\half}V_1=-2\left[ \frac{d}{d\theta}(\xi^{-\half}\del_1)+\frac{1}{\xi} \int^{\theta}
\xi^{-3/2}(\theta') \del_1(\theta') \ d\theta'\right].$$ Hence
\BEA
\int^{\theta} V_1(\theta')\del_1(\theta')\ d\theta' &=& \int^{\theta} (\xi^{\half}V_1)(\theta')
(\xi^{-\half}\del_1)(\theta') \ d\theta' \nonumber\\
&=& - \left[ \left( \xi^{-\half}(\theta)\del_1(\theta) \right)^2+ \left( \int^{\theta} \xi^{-3/2}(\theta')\del_1(\theta')
\ d\theta' \right)^2 \right]
\EEA
and the integral over a period is zero. \hfill \eop

The following covariance result is an extension of Theorem 3.1.6.

{\bf Theorem 4.4}

{\em
Let $D\in {\cal S}_{\le 2,gen}^{aff}$ be of generic type, with associated invariant $(\xi=\xi(D),\del_1=\del_1(D),
\del_2=\del_2(D))$. Then:
\begin{enumerate}
\item
\BEQ {\cal EL}(D):=\half\left[ \frac{1}{\xi}(1+\frac{1}{4} \dot{\xi}^2)x^2- \xi \partial_x^2
+\frac{\II}{2} \dot{\xi} (x\partial_x+\partial_x x)+\left(-2\del_1 (-\II \partial_x)+(V_1 \xi+2\dot{\del_1})x\right)
+2(\del_2+\half V_0 \xi) \right] \EEQ
is an invariant for the Schr\"odinger operator $D$.
\item
Let $(\phi;(a,b))\in \SV$ and $g: (\theta,x)\to (\theta',x')=(\phi(\theta),x\sqrt{\dot{\phi}(\theta)}-a(\theta))$ be the
associated coordinate change. Then
\BEQ \pi_{1/4}(\phi;(a,b)) {\cal EL}(D) \pi_{1/4}(\phi;(a,b))^{-1} = \widetilde{\cal EL}(D) \label{transfELxideleps} \EEQ
where $\widetilde{\cal EL}(D)$ is obtained by applying the transformation $g$ to the coordinates, 
changing the potentials $V_0$ and $V_1$ by the $\sigma_{1/4}$-action of SV, and transforming
the invariant  as follows:
\BEQ \tilde{\xi}=\phi' \ . \ \xi\circ \phi^{-1}; \EEQ
\BEQ \tilde{\del_1}=\phi'^{\half}\ .\ \del_1\circ \phi^{-1} + (\tilde{\xi} \dot{a}-\half a\dot{\tilde{\xi}});\EEQ
\BEQ \tilde{\del_2}=\del_2\circ\phi^{-1} + (\del_1 \dot{a}-a\dot{\del_1}) + \tilde{\xi} \dot{b}+(\tilde{\xi}(\dot{a}^2-a\ddot{a})
-\dot{\tilde{\xi}}a\dot{a}-\ddot{\tilde{\xi}}a^2).  \EEQ
Furthermore, $(\tilde{\xi},\tilde{\del_1},\tilde{\del_2})$ is the invariant associated to $\sigma_{1/4}(D)$.
\end{enumerate}
}

{\bf Proof.}

\begin{enumerate}
\item
Look for an invariant of the form
\BEQ \half\left[ a(\theta) x^2-b(\theta)\partial_x^2-\II c(\theta) (x\partial_x+\partial_x x)+ d(\theta) (-\II \partial_x)+e(\theta) x+f(\theta) \right] \label{abcdef} \EEQ
and solve in $a,b,c,d,e,f$. One obtains  the following constraints:
\BEQ \dot{a}=2V_2 c,\quad \dot{b}=-2c, \quad \dot{c}=-a+V_2 b \EEQ
-- whose general solution is in Proposition 3.1.4 above, namely, $a=\frac{1}{\xi} (1+\frac{1}{4} \dot{\xi}^2)$,
$b=\xi$, $c=-\half \dot{\xi}$ -- and the set of following equations:
\BEQ \dot{d}=V_1 b-e,\quad \dot{e}=V_1 c+V_2 d,\quad \dot{f}=\half d V_1 \EEQ
which implies the compatibility condition
$$\ddot{d}+V_2 d=\dot{V}_1 \xi+\frac{3}{2} V_1 \dot{\xi}.$$

\item

Since (assuming $I_u(\xi)=2$ is fixed) there is a unique invariant for operators of generic type, one necessarily has
\BEQ {\cal L}_{\tilde{\xi}}+{\cal Y}_{\tilde{\del_1}}+{\cal M}_{\tilde{\del_2}}=Ad(\phi;(a,b)).({\cal L}_{\xi}
+{\cal Y}_{\del_1}+{\cal M}_{\del_2})\EEQ
which gives the above formulas for $(\tilde{\xi},\tilde{\del_1},\tilde{\del_2})$.

There remains to check for equation (\ref{transfELxideleps}). Consider first the covariance under a  time-reparametrization $\phi$. It has already been proved for the quadratic part of the Ermakov-Lewis operator, see Theorem 3.1.6.
The linear part $-2(-\II \del_1 \partial_x+(V_1 \xi-\dot{\del_1})x)$ transforms covariantly under $\phi$ since
(see proof of Theorem 3.1.6)
\BEQ \tilde{V}_1 \tilde{\xi}-\frac{d\tilde{\del_1}}{d\theta'}=\phi'^{-\half} (V_1.\xi-\dot{\del_1}-\half \frac{\ddot{\phi}}{\dot{\phi}}\del_1), \quad -\II \tilde{\del_1}\partial_{x'}=-\II \del_1 \partial_x \EEQ
and
\BEA
&&  \left( \pi_{1/4}(\phi) (-\II \del_1(\theta)\partial_x+(V_1 \xi-\dot{\del}_1)x)\pi_{1/4}(\phi^{-1})\right)\psi =\nonumber\\
&& =\half \dot{\phi}^{-1/4} e^{\frac{\II}{4} \frac{\ddot{\phi}}{\dot{\phi}}x^2} (-\II \del_1 \partial_x+
(V_1 \xi-\dot{\del}_1)x)\dot{\phi}^{1/4} e^{-\frac{\II}{4} \frac{\ddot{\phi}}{\dot{\phi}}x^2} \psi(\phi(\theta),
x\sqrt{\dot{\phi}(\theta)})  \nonumber\\
&&= (-\II \tilde{\del_1} \partial_{x'} +(\tilde{V}_1 \tilde{\xi}-\tilde{\del_1})x')\psi(\theta',x').
\EEA
As for the zero-order term $-\half(\del_2+\half V_0 \xi)$, it is obviously invariant under the conjugate
action of $\pi_{1/4}(\phi)$. Since $\tilde{V}_0 \tilde{\xi}=(V_0 \xi)\circ\phi^{-1}$, this implies
also $\tilde{\del_2}=\del_2\circ\phi^{-1}$.

Consider now the covariance under an infinitesimal nilpotent transformation ${\cal Y}_{f_1}+{\cal M}_{f_2}$.
One has
\BEA
&&  [a\partial_x+\II \dot{a}x+b,{\cal EL}(\xi,\del_1,\del_2)]= \nonumber\\
&& \half [a\partial_x+\II \dot{a}x,\frac{1}{\xi}(1+\frac{1}{4}\dot{\xi}^2)x^2-\xi\partial_x^2+\II \dot{\xi}
x\partial_x-2(\del_1(-\II \partial_x)+(V_1 \xi-\dot{\del_1})x) ] \nonumber\\
&& =\half \left\{ \left(\frac{2a}{\xi}(1+\frac{1}{4}\dot{\xi}^2)+\dot{a}\dot{\xi}\right)x-(a\dot{\xi}+2\dot{a}
\xi)(-\II \partial_x)-2a(V_1 \xi-\dot{\del_1})+2\dot{a}\del_1 \right\},
\EEA
to be compared with the infinitesimal change of ${\cal EL}$ under the transformation $x\to x+\eps a$,
$\del_1\to \del_1+\eps (\xi \dot{a}-\half a\dot{\xi})$, $\del_2\to \del_2+\eps((\del_1 \dot{a}-a\dot{\del_1})+\xi \dot{b})$,
$V_1\to V_1-2\eps(\ddot{a}+V_2 a)$. This is a straightforward computation, which requires the use of 
the equation defining $\xi$, namely, $\ddot{\xi}=\frac{1}{\xi}(1+\frac{1}{4}\dot{\xi}^2)-2V_2\xi$.

\end{enumerate}

\hfill \eop

\vskip 3 cm

Using the parametrization of ${\cal S}_{\le 2,gen}^{aff}$ by the vector invariant $(\xi,\del_1,\del_2)$,
one can easily define a natural symplectic structure on a linear space $\Omega$ and a hamiltonian action of $SV$ on $\Omega$
reproducing the SV-action on ${\cal S}_{\le 2,gen}^{aff}$. 

{\bf Definition 4.5}

{\em 
Let  $\Omega\simeq C^{\infty}(\R/2\pi\Z,\R^4)$ be the linear  manifold consisting of all $2\pi$-periodic vector-valued
$C^{\infty}$ functions $\vec{X}(\tau):=(p,q,E,t)(\tau)$, $\tau\in\R/2\pi\Z$ with singular Poisson structure defined by
\BEQ \{ p(\tau),q(\tau')\}=\delta(\tau-\tau'), \quad \{ E(\tau),t(\tau')\}=\delta(\tau-\tau') \EEQ
}

See for instance \cite{GuiRog}, chap. X for some remarks on distribution-valued singular Poisson structures on infinite-dimensional
spaces. The energy $E$ is canonically conjugate to $t$, which allows us to consider generalized canonical transformations for which $t$ is a coordinate. This usual trick for hamiltonian systems with time-dependent Hamiltonians can for instance be found in
\cite{Gol59}. Hamitonian vector fields ${\cal X}_H$, for $H=H(p,q,E,t)$, acts  separately on each fiber $\tau=$constant, namely,
\BEQ ({\cal X}_H f)(\tau):=\left\{ (\partial_p H \partial_q -\partial_q H \partial_p+\partial_E H \partial_p-\partial_t H
\partial_E)f \right\}(\tau).\EEQ

{\bf Definition 4.6}

{\em
Let $(\xi,\del_1,\del_2)$ be a triple of $2\pi$-periodic functions. Define $\Phi:=\Phi(\xi,\del_1,\del_2)$ to be the following
functional on $\Omega$,
\BEQ \langle \Phi,\vec{X}\rangle=\oint \left\{ \xi(t(\tau))E(\tau)+\half \dot{\xi}(t(\tau))p(\tau)q(\tau)+\del_1(t(\tau))
p(\tau)-\dot{\del}_1(t(\tau))q(\tau) +\del_2(t(\tau)) \right\} \ d\tau.\EEQ
}

{\bf Theorem 4.7}

{\em
Represent ${\cal L}_f+{\cal Y}_g+{\cal M}_h\in \sv$ by the hamiltonian vector field ${\cal X}_{H(f,g,h)}$ associated to
\BEQ H(f,g,h):=-(f(t)E+\half \dot{f}(t) pq+\frac{1}{4} \ddot{f} q^2)-(g(t)p+\dot{g}(t) q)-h(t).\EEQ
Then the action of ${\cal X}_H$ on the functional $\Phi(\xi,\del_1,\del_2)$ coincides with that given in Theorem 4.4.
}

{\bf Proof.}

Observe that the map from $\sv$ to the Lie algebra of vector fields on $\Omega$ given by 
${\cal L}_f+{\cal Y}_g+{\cal M}_h\to {\cal X}_{H(f,g,h)}$ is a Lie algebra homomorphism. The vector field ${\cal X}_H$
is given explicitly by
\BEA
{\cal X}_{H(f,g,h)}&=&-\left[ \half \dot{f}(t)(q\partial_q-p\partial_p)+f(t)\partial_t-\half \ddot{f}(t)q\partial_p\right]
-\left[g(t)\partial_q-\dot{g}(t)\partial_p\right] \nonumber\\
&+& \left[ (\half \ddot{f}(t) pq+\dot{f}(t) E+\frac{1}{4} f'''(t)q^2)+(\dot{g}p+\ddot{g}q)+\dot{h}(t)\right] \partial_E.
\nonumber\\
\EEA
The rest is a straightforward computation.
\hfill \eop

\vskip 3 cm

Let us conclude this section by computing the monodromy for 'resonant' operators of type (i)bis and (iii)bis.

Consider any resonant operator $D$. The associated classical monodromy is unipotent. We choose $\xi\in \Stab_{V_2}$
to be purely imaginary, $\xi:=\II\eta$ as before (see section 3.2). A generalized Ermarkov-Lewis
invariant may then be defined as
\BEQ {\cal EL}(D)=\frac{1}{2\xi} \left[ (\II \xi\partial_x+\half \dot{\xi}x)^2\right] +\frac{\II}{2}
\left[ d(-\II\partial_x)+ex+f\right],\EEQ
where $d,e,f$ are defined as in Theorem 4.4 but with $\xi$ replaced by $\eta$ (see equation (\ref{abcdef}) for notations). Hence
\BEQ \frac{ {\cal EL}(D)-\II k}{\xi}=-\half\left[ (\partial_x-\frac{\II}{2}\frac{\dot{\eta}}{\eta}x)^2-
\frac{d}{\eta}(-\II\partial_x)-\frac{e}{\eta} x-\frac{f}{\eta} \right]-\frac{k}{\eta}.\EEQ
Suppose ${\cal EL}(D)\psi_k=\II k\psi_k$ and set
 \BEQ \tilde{\psi}_k=\exp \left( -\frac{\II}{4} \frac{\dot{\eta}}{\eta}
x^2+\frac{\II}{2} \frac{d}{\eta} x\right) \psi_k.\EEQ
 Then a simple calculation gives
\BEQ \left[ \partial_x^2-\left(\half d\frac{\dot{\eta}}{\eta^2}+\frac{e}{\eta}\right)x+\frac{-f+2k}{\eta}+\frac{1}{4}
\left( \frac{d}{\eta}\right)^2 \right] \tilde{\psi}_k=0 \label{transfELbis} \EEQ

If $D$ is of type (i)bis, then $d,e,f$ (easy to obtain from Theorem 4.4 and the isotropy algebra given in section 2)
satisfy $\half d\frac{\dot{\eta}}{\eta^2}+\frac{e}{\eta}=0$ identically, so the model operator is (up to a constant)
the Laplacian as for case (iii). Then the monodromy can be computed along the same lines as in section 3.6, with
a time-independent shift in $k$ due to the function $-f+\frac{1}{4} \frac{d^2}{\eta}=\frac{3C^2}{128n^2}.$

{\bf Lemma 4.8}

{\em 
Let $D=-2\II \partial_{\theta}-\partial_x^2+n^2 x^2+C\cos n(\theta-\sigma/2)x+\gamma$ be a Schr\"odinger
operator of type (i)bis. Set
\BEQ \psi_{k,\pm}(\theta,x)=e^{\frac{\II}{4} \frac{\dot{\eta}}{\eta} x^2-\frac{\II}{2} \frac{d}{\eta}x}
\ .\ e^{\pm\II \sqrt{\frac{2k'}{\eta}} x},\EEQ
with $d=-\frac{C}{4n}\sin 3n(\theta-\sigma/2)$, $\eta=1-\cos(2n\theta-\sigma)$, $k'=k+3\left(\frac{C}{16n}\right)^2$. Then 
\BEQ D\psi_{k,\pm}=\left(\frac{2k'}{\eta} +\frac{1}{4} \left(\frac{d}{\eta}\right)^2\mp d\eta^{-3/2}
\sqrt{2k'}-\frac{\II}{2}\frac{\dot{\eta}}{\eta}+\gamma\right) \psi_{k,\pm}.\EEQ
}

{\bf Proof.} Tedious computations. \hfill \eop

Apart from the time-periodic shift $\frac{1}{4} \left(\frac{d}{\eta}\right)^2=\left(\frac{C}{16n}\right)^2
\frac{\sin^2 3n(\theta-\sigma/2)}{\sin^4 n(\theta-\sigma/2)}$ (which is integrable on the contour $\Gamma$)
and the time-independent shift in $k$, one is left once again with a phase proportional to 
$k/\eta$ (note that the term in $d\eta^{-3/2} \sqrt{2k'}$ is irrelevant since $\int_0^{2\pi}
(d\eta^{-3/2})(\theta)\ d\theta=0$ by Lemma 4.3; recall $d=-2\del_1$ by Theorem 4.4).

Hence one obtains:

{\bf Theorem 4.9}

{\em

Let $\psi(0)\in L^2(\R)$, with decomposition
\BEQ \psi(0,x):= \int_{\R_+} \bar{c}_+(k) \psi_{k,+}(0,x) \ dk +\int_{\R_+} \bar{c}_-(k)\psi_{k,-}(0,x)\ dk.\EEQ
Then the solution of the type (i)bis  Schr\"odinger equation
\BEQ (-2\II\partial_{\theta}+\partial_x^2+n^2 x^2+C\cos (n\theta-\sigma/2).x+\gamma)\psi=0 \EEQ
 with initial state $\psi(0)$ is given at time $\theta=2\pi$ by
\BEQ \psi(2\pi,x)= \int_{\R_+} \bar{c}_+(k) e^{ k' T-\II\pi \tilde{\gamma}} \psi_{k,+}(0,x)\ dk
+  \int_{\R_+} \bar{c}_-(k) e^{k'T-\II\pi\tilde{\gamma}} \psi_{k,-}(0,x)\ dk,\EEQ
where $k'=k+3\left(\frac{C}{16n}\right)^2$, $T=\int_0^{2\pi} \frac{du}{\xi(u)}$ ($T$ is purely imaginary) and
$$\tilde{\gamma}=\gamma+\frac{1}{4} \int_{\Gamma} \left(\frac{d}{\eta}\right)^2(\theta) d\theta.$$
The associated monodromy operator in ${\cal B}(L^2(\R),L^2(\R))$ is unitarily equivalent to the unitary
operator on $L^2(\R_+)$ given by the multiplication by the function $k\to e^{kT-\II\pi\tilde{\gamma}}$.
}

\vskip 1 cm

Suppose now $D$ is of type (iii)bis. Then the $x$-coefficient in the transformed Ermakov-Lewis operator
(\ref{transfELbis})  does not
vanish, so one must take for 'model operator' $-\partial_x^2+x$, whose eigenfunctions are related to the Airy function.
The solution of the monodromy problem will be given by a series of lemmas. In the sequel, $\eta=(1+\sin n\theta)(1+\alpha\sin n\theta)$ is the (real-valued and non-negative) invariant, and $\eta^{1/2}=(1+\alpha\sin n\theta)^{1/2} \cos(\frac{\pi}{4}-n\frac{\theta}{2})x$ is the smooth square-root of $\eta$ chosen in section 2.

{\bf Lemma 4.10}

{\em
Let $Ai$ be the entire function, solution of the Airy differential equation $(-\partial_x^2+x)Ai(x)=0$,  defined on the real line as
\BEQ Ai(x)=\frac{1}{\pi} \int_0^{\infty} \cos\left( \frac{t^3}{3}+xt \right) \ dt.\EEQ
It is (up to a constant) the only solution of the Airy differential equation which do not increase exponentially
on $\R_+$. The functions $f_k(x):=Ai(x-k)$, $k\in\R$ define (up to a coefficient) a complete orthonormal
system of generalized eigenfunctions   of the self-adjoint closure of the Airy operator $-\partial_x^2+x$ with core
$C_0^{\infty}(\R)\subset L^2(\R)$. 
}

{\bf Proof.} Easy by using a Fourier transform. \hfill \eop

{\bf Lemma 4.11} 

{\em
The $x$-coefficient in the transformed Ermakov-Lewis invariant (\ref{transfELbis}) reads
\BEQ \half d\frac{\dot{\eta}}{\eta^2}+\frac{e}{\eta}=-C_{\alpha} \eta^{-3/2}  \label{lemma4.10} \EEQ
where $C_{\alpha}=(1-\alpha)(1+\alpha/2)\sqrt{1-\alpha^2}.$
}

{\bf Proof.}

Computations similar to that of Lemma 4.3 (with the simple difference that $\xi \ddot{\xi}-\half \dot{\xi}^2+2V_2\xi^2=0$
here) yield
\BEQ  \frac{d}{d\theta} (\eta^{-\half}d)=\eta -\frac{C_{\alpha}}{\eta} \EEQ
where $C_{\alpha}$ is some constant which must be chosen in order that the right-hand side be $2\pi$-periodic. Note that
the singularities in the above equation are only apparent; one may avoid them altogether by using a contour $\Gamma$
in the upper-half plane as in section 3.2. 
Since $\int_0^{2\pi} \eta=2\pi(1+\alpha/2)$ and $\int_{\Gamma} \frac{d\theta'}{\eta(\theta')}=-\frac{2\pi}{(1-\alpha)
\sqrt{1-\alpha^2}}$ (see Proposition 2.2.3), this gives $C_{\alpha}=(1-\alpha)(1+\alpha/2)\sqrt{1-\alpha^2}$.
Then a straightforward computation yields formula  (\ref{lemma4.10}). \hfill \eop

{\bf Lemma 4.12}

{\em
Set 
\BEQ \psi_k(\theta,x)=\exp\left( \frac{\II}{4} \frac{\dot{\eta}}{\eta}x^2-\frac{\II}{2} \frac{d}{\eta}x\right)
 \ . \eta^{-\half} \ Ai\left(xC_{\alpha}^{1/3} \eta^{-\half}-C_{\alpha}^{-2/3}(-f+2k+\frac{1}{4} \frac{d^2}{\eta} )\right)
\EEQ
Then 
\BEQ D\psi_k(\theta,x)=\left(\frac{2k}{\eta}+\left( \frac{\II}{2} \frac{\dot{\eta}}{\eta}+\half
\left(\frac{d}{\eta}\right)^2 - \frac{f}{\eta} \right) \right) \psi_k(\theta,x). \label{lemma4.11} \EEQ
}

{\bf Proof.}

The $\psi_k$ are obtained as in section 3.5 (monodromy of hyperbolic operators) by taking a complete orthonormal
system of generalized eigenfunctions $\tilde{\psi}_k$ for the transformed Ermakov-Lewis invariant (\ref{transfELbis}) and
going back to the functions $\psi_k$. Then (\ref{lemma4.11}) is proved by a direct tedious computation. \hfill \eop

One may now conclude:

{\bf Lemma 4.13}

{\em

Let $\psi(0)\in L^2(\R)$, with decomposition
\BEQ \psi(0,x):= \int_{\R} \bar{c}(k) \psi_k(0,x) \ dk.\EEQ
Then the solution of the type (iii)bis  Schr\"odinger equation
\BEQ \left(-2\II\partial_{\theta}+\partial_x^2+v_{n,\alpha}  x^2+C(1+\alpha\sin n\theta)^{\half}
\cos (\frac{\pi}{4}-n\frac{\theta}{2}).x+\gamma \right)\psi=0 \EEQ
 with initial state $\psi(0)$ is given at time $\theta=2\pi$ by
\BEQ \psi(2\pi,x)= \int_{\R} \bar{c}(k) e^{ k T-\II\pi \tilde{\gamma}}  \psi_k(0,x) \ dk, \EEQ
where $T=\int_0^{2\pi} \frac{du}{\xi(u)}$ ($T$ is purely imaginary) and
$$\tilde{\gamma}=\gamma+\int_{\Gamma} \left( -\frac{f}{\eta}+\frac{1}{2}\left(\frac{d}{\eta}\right)^2\right)(\theta)\ d\theta.$$
The associated monodromy operator in ${\cal B}(L^2(\R),L^2(\R))$ is unitarily equivalent to the unitary
operator on $L^2(\R)$ given by the multiplication by the function $k\to e^{kT-\II\pi\tilde{\gamma}}$.
}

{\em Acknowledgements.} We wish to thank A. Joye for a useful discussion in Cergy.

\end{document}